\pdfoutput=1
\newcommand*{\ATLASLATEXPATH}{}
\documentclass[UKenglish,texlive=2016,cernpreprint]{\ATLASLATEXPATH atlasdoc}
\usepackage{\ATLASLATEXPATH atlaspackage}
\usepackage{\ATLASLATEXPATH atlasbiblatex}
\usepackage{\ATLASLATEXPATH atlasphysics}

\usepackage{HIGG-2017-05-PAPER-defs}

\usepackage{amsmath,amssymb}
\usepackage{tikz}
\usepackage{multirow}

\AtlasTitle{Search for the Higgs boson produced in association with a
vector boson and decaying into two spin-zero particles in the
$H \rightarrow aa \rightarrow 4b$ channel in
$\boldsymbol{pp}$ collisions at $\sqrt{\textbf{s}}$~$\boldsymbol{=}$~13~\TeV\ 
with the ATLAS detector}

\author{The ATLAS Collaboration}
\AtlasRefCode{HIGG-2017-05}
\PreprintIdNumber{CERN-EP-2018-128}
\AtlasJournal{JHEP}
\AtlasJournalRef{JHEP 10 (2018) 031}
\AtlasDOI{10.1007/JHEP10(2018)031}

\AtlasAbstract{
A search for exotic decays of the Higgs boson into a pair of spin-zero
particles, $H \rightarrow aa$, where the $a$-boson decays into $b$-quarks
promptly or with a mean proper lifetime $c\tau_a$ up to 6~mm
and has a mass in the range of $20$--$60~{\GeV}$, is presented. 
The search is performed in events where the Higgs boson is produced in
association with a $\Wboson$ or $\Zboson$ boson, giving rise to a
signature of one or two charged leptons (electrons or muons) and multiple jets
from $b$-quark decays. The analysis is based on the dataset of
proton--proton collisions at $\sqrt{s} = 13~{\TeV}$ recorded in
2015 and 2016 by the ATLAS detector at the CERN Large Hadron Collider,
corresponding to an integrated luminosity of $36.1~{\ifb}$. No
significant excess of events above the Standard Model background prediction is 
observed, and $95\%$ confidence-level upper limits are derived for the 
production cross-sections for $pp \to WH$, $ZH$ and their combination, 
times the branching ratio of the decay chain $H \to aa \to 4b$. For
$a$-bosons which decay promptly, the upper limit on the combination of
cross-sections for $WH$ and $ZH$ times the branching ratio of $H \to aa
\to 4b$ ranges from 3.0~pb for $m_a= 20~{\GeV}$ to 1.3~pb for $m_a =
60~{\GeV}$, assuming that the ratio of $WH$ to $ZH$ cross-sections
follows the Standard Model prediction. For $a$-bosons with longer proper lifetimes, the most
stringent limits are 1.8~pb and 0.68~pb, respectively, at $c\tau_a \sim 0.4$~mm.
}

\hypersetup{pdftitle={ATLAS document},pdfauthor={The ATLAS Collaboration}}

\begin{document}

\maketitle

\section{Introduction}
\label{sec:intro}

The discovery of the Higgs boson by the ATLAS and CMS 
collaborations~\cite{ATLAS_Higgs,CMS_Higgs} at the Large Hadron
Collider (LHC) has been a major achievement for the Standard Model (SM). A comprehensive
programme to explore the properties of this particle is underway,
including measurements of the branching ratios to SM particles and
searches for decays into ``exotic'' or non-SM particles. Exotic Higgs
boson decays are a powerful probe for physics
beyond the SM (BSM). The Higgs boson has a very narrow decay width, so even
a small coupling to a non-SM particle could open up a sizeable decay
mode. Measurements at the LHC are in agreement with SM predictions,
constraining the non-SM branching ratio of the Higgs
boson to less than approximately $30\%$ at $95\%$ confidence level
(CL) using the 7~and~8~\TeV\
datasets~\cite{Aad:2015gba,Khachatryan:2014jba,Khachatryan:2016vau}. Despite
this experimental triumph, there is still ample room for exotic Higgs boson
decays compatible with observations to date.

The Higgs boson has been proposed as a possible ``portal'' for
hidden-sector particles to interact with SM 
particles~\cite{Strassler:2006im,Schabinger:2005ei,Patt:2006fw}. 
Exotic decays in particular are predicted by many BSM
theories~\cite{Curtin:2013fra}, including those with an extended Higgs sector
such as the Next-to-Minimal Supersymmetric Standard Model 
(NMSSM)~\cite{Dobrescu:2000yn,Ellwanger:2003jt,Dermisek:2005ar,Chang:2008cw,Morrissey:2008gm},
models with a first-order electroweak phase
transition~\cite{Profumo:2007wc,Blinov:2015sna}, 
models with neutral naturalness~\cite{Burdman:2006tz,Craig:2015pha,Curtin:2015fna} and models of dark
matter~\cite{Silveira:1985rk,Pospelov:2007mp,Draper:2010ew,Ipek:2014gua,Martin:2014sxa}. 

The decay of the Higgs boson into a pair of spin-zero particles $a$, which in turn decay into a pair of
SM particles, arises in several scenarios of new physics~\cite{Curtin:2013fra}. 
In particular, if the $a$-boson mixes
with the Higgs boson and inherits its Yukawa couplings, decays of the
type $a\to \bbbar$ are expected to be dominant for $m_a > 2m_b$. 
Models of this type have been used to explain the observations of a 
gamma-ray excess from the galactic centre by the Fermi Large Area Telescope
(FermiLAT)~\cite{TheFermi-LAT:2015kwa,Boehm:2014hva}.
In models of neutral naturalness, the $a$-boson could have 
mean proper lifetimes ($c\tau_a$) ranging from about $10~\mu$m 
to $\mathcal{O}$(km)~\cite{Curtin:2015fna}. Lifetimes smaller
than $10~\mu$m are referred to as ``prompt''. 

This paper considers the decay mode $H \to aa$ with the subsequent
decay $a\to \bbbar$, building on the previous work of
ref.~\cite{Aaboud:2016oyb}, in which a similar analysis was
reported with a subset of the data considered here. The previous
result set an upper limit on the production cross-section $\sigma(WH)$ 
times the branching ratio for $H \to aa \to 4b$ ranging from $6.2$~pb for
an $a$-boson mass of $m_a= 20~{\GeV}$ to 1.5~pb for $m_a =
60~{\GeV}$, compared with the SM cross-section $\sigma_{\mathrm{SM}}(WH) = 1.37$~pb.
This paper includes ten times more data, adds the $ZH$ channel and an
improved analysis technique.

\begin{figure*}[htbp]
\begin{center}
        \subfloat[]{\includegraphics[width=0.5\textwidth]{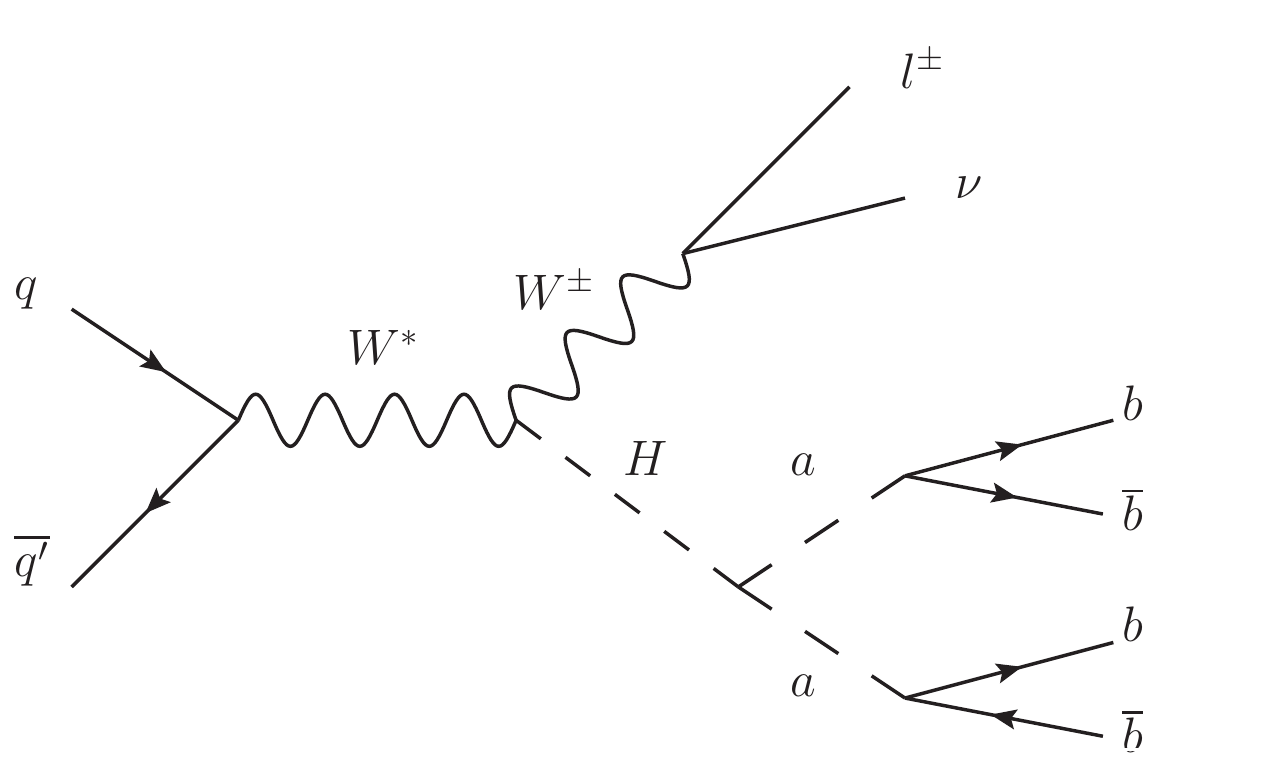}}%
        \subfloat[]{\includegraphics[width=0.5\textwidth]{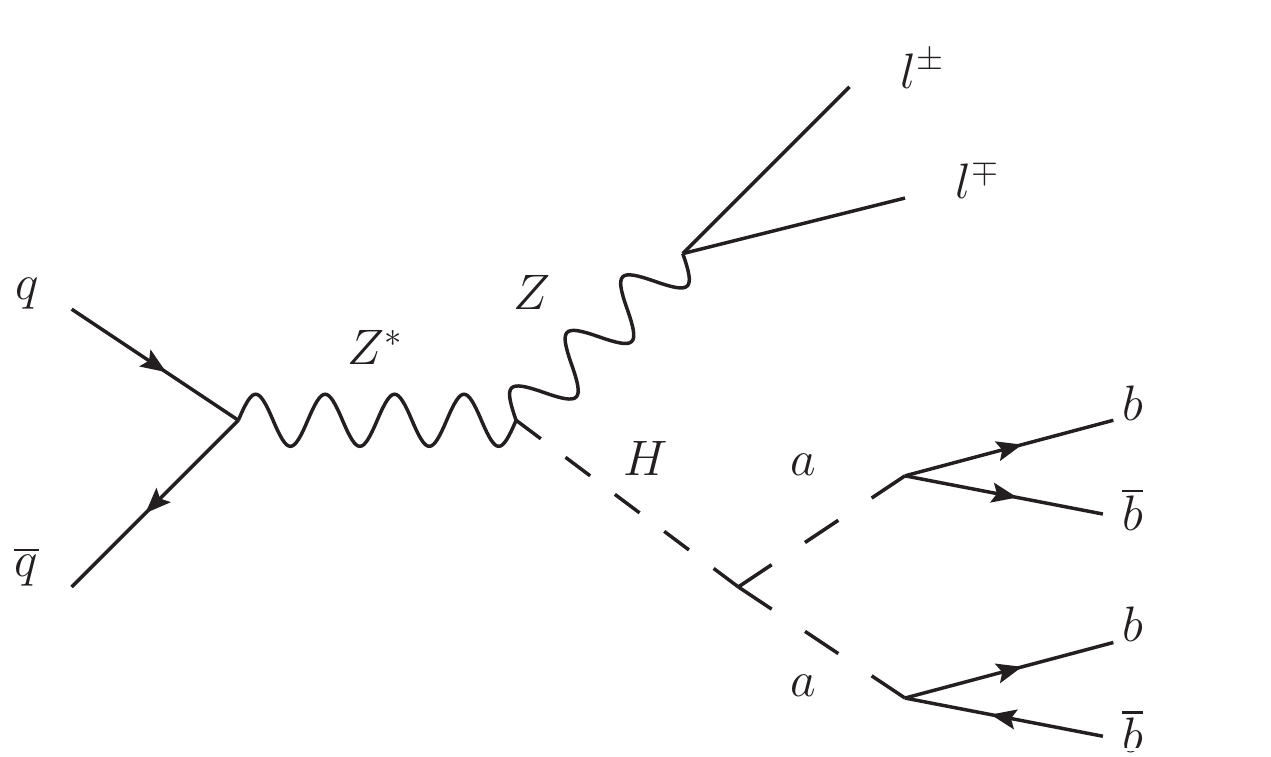}}\\
        \caption{Representative tree-level Feynman diagrams for the
          (a) $WH$ and (b) $ZH$ production processes with the subsequent decays
          $W\to\ell\nu$, $Z\to\ell\ell$ ($\ell=e,\mu$) and $H \to  aa \to 4b$.}
    \label{fig:feynmandiagrams}
\end{center}
\end{figure*}

This search focuses on the $WH$ and $ZH$ processes, with $W\to\ell\nu$, $Z\to\ell\ell$
($\ell=e,\mu$) and $H \to  aa \to 4b$, as shown in figure~\ref{fig:feynmandiagrams}.
The $a$-boson can be either a scalar or a pseudoscalar under parity
transformations, since the decay mode considered in this search is not
sensitive to the difference in coupling. The $a$-boson signals
considered have masses in the range $20~{\GeV} \leq m_a \leq 60~{\GeV}$ and mean
proper lifetimes, $c\tau_a$, up to 6~mm.

The resulting signature has a single lepton or two leptons accompanied by a high 
multiplicity of jets originating from $b$-quarks ($b$-jets). Since
four $b$-jets are produced from the decay of the Higgs boson, they tend to 
have a low transverse momentum ($\pt$) compared with $m_H$ and can 
overlap, especially for light $a$-bosons. Events with one or two
electrons or muons, including those produced via leptonically 
decaying $\tau$-leptons, are considered. The $WH$ and $ZH$ processes 
are chosen for this search because the presence of at least one
charged lepton in the final state provides a powerful 
signature for triggering and for suppressing background from the high cross-section
strong-interaction production of four $b$-jets. 

Several kinematic variables, including the reconstructed masses in 
the decay $H\to aa \to 4b$, are combined to identify signal events. The 
background estimation techniques, systematic uncertainties and 
statistical treatment closely follow those used in other ATLAS
searches with similar signatures~\cite{Aaboud:2017rss,Aad:2015gra,Aad:2015kqa,Aad:2015pja,Aad:2015typ}.

\section{ATLAS detector}
\label{sec:detector}

The ATLAS experiment~\cite{PERF-2007-01} is a multipurpose particle
physics detector with forward--backward symmetric cylindrical geometry
and nearly 4$\pi$ coverage in solid angle.\footnote{ATLAS uses a
  right-handed coordinate system with its origin at the nominal
  interaction point (IP) in the centre of the detector and the
  $z$-axis along the beam pipe. The $x$-axis points from the IP to the
  centre of the LHC ring, and the $y$-axis points upward. Cylindrical
  coordinates $(r,\phi)$ are used in the transverse plane, $\phi$
  being the azimuthal angle around the $z$-axis. The pseudorapidity is
  defined in terms of the polar angle $\theta$ as
  $\eta=-\ln\tan(\theta/2)$. The transverse momentum \pt~and other transverse variables
  are defined by projecting these variables into the $x$--$y$ plane, and
  the transverse energy \et is defined as $\sqrt{m^2+ \pt^2}$, where
  $m$ represents the mass of a considered
  object. The distance in the pseudorapidity--azimuthal-angle space
  is defined as $\Delta R = \sqrt{(\Delta \eta)^2 + (\Delta
    \phi)^2}$.}  The interaction point is surrounded by
an inner detector (ID) tracking system, a calorimeter system, and a muon spectrometer
(MS). The ID covers $|\eta| < 2.5$ and consists of a silicon pixel
detector, a silicon microstrip detector, and a transition radiation
tracker. The ID includes the insertable B-layer~\cite{IBL},
a pixel layer close to the interaction point, which
provides high-resolution measurements at small radius
to improve the tracking performance. A thin superconducting solenoid
surrounds the ID and provides a 2 T axial magnetic field. The calorimeter system features a
high-granularity lead/liquid-argon sampling calorimeter that
measures the energy and the position of electromagnetic showers within
$|\eta| < 4.9$. Liquid-argon sampling calorimeters are also used to measure
hadronic showers in the endcap ($1.5 < |\eta| < 3.2$) and forward
($3.1 < |\eta| < 4.9$) regions, while a steel/scintillator tile
calorimeter measures hadronic showers in the central region ($|\eta| <
1.7$). The MS surrounds the calorimeters and consists of three large 
superconducting air-core toroid magnets, each with eight coils, a
system of precision tracking chambers ($|\eta| < 2.7$), and fast
trigger chambers ($|\eta| < 2.4$). For Run 2, the ATLAS detector has a 
two-level trigger system. The first-level trigger is
implemented in hardware and uses a subset of the detector information
to reduce the rate of accepted events to 100 kHz. This is followed by 
the sofware-based high-level trigger that reduces
the rate of recorded events to 1 kHz.

\section{Event and object selection}
\label{sec:selection}

Events are selected from proton--proton ($pp$) collisions collected by the ATLAS 
detector at the LHC at $\sqrt{s}=13~{\TeV}$ in 2015 and 2016. 
The dataset corresponds to an integrated luminosity of $3.2 \pm
0.1$~fb$^{-1}$ recorded in 2015 and $32.9 \pm 0.7$ fb$^{-1}$ recorded in 2016, 
for a total of $36.1 \pm 0.8$ fb$^{-1}$~\cite{DAPR-2013-01}.
The data used for this search were collected using the single-electron
or single-muon triggers with the lowest $\pt$ thresholds 
available, 20 (26)~\GeV\ for muons and 24 (26)~\GeV\ for
electrons in 2015 (2016)~\cite{TriggerRun2}. 

Electrons are reconstructed from energy deposits (clusters) in the
electromagnetic calorimeter matched to tracks in the
ID~\cite{Aad:2014fxa} and are required to have $\pt>15~{\GeV}$ and
$|\eta|<2.47$. Candidates in the transition region between the barrel and 
endcap calorimeters, $1.37<|\eta|<1.52$, are excluded. Electrons must
satisfy the ``tight'' identification criterion based on a likelihood
discriminant~\cite{ElectronID2016}. Muons are
reconstructed by combining matching tracks in the ID and the MS, and
are required to have $\pt>10~{\GeV}$ and $|\eta|<2.4$. Muon candidates 
must satisfy the ``medium'' identification criterion~\cite{PERF-2015-10}.

In order to distinguish leptons produced in the decays of $\Wboson$
and $\Zboson$ bosons from those produced in the decays of
heavy-flavour hadrons, all lepton candidates are required to originate
from the primary interaction vertex, chosen as the vertex with the
highest sum of the $\pt^2$ of its associated tracks. 
Furthermore, since lepton candidates arising from background sources, 
such as decays of hadrons, are typically embedded in jets, all lepton
candidates are required to be isolated from other particles in the
event. This is achieved by imposing a maximal allowed value on the
energy deposited in the calorimeter and/or the momentum of
ID tracks within a cone around the direction of the lepton candidate,
according to the gradient isolation criteria~\cite{ElectronID2016,PERF-2015-10}. 

Jets are reconstructed from three-dimensional topological energy
clusters~\cite{PERF-2014-07} in the calorimeter 
using the \antikt\ jet algorithm~\cite{Cacciari:2008gp} implemented in
the FastJet package~\cite{Cacciari:2011ma} with a radius parameter of
0.4.  Jets are calibrated using energy- and $\eta$-dependent
corrections~\cite{PERF-2016-04} and are required to have $\pt
>20~{\GeV}$ and $|\eta|<2.5$.
Events containing jets arising from non-collision sources or detector
noise are removed~\cite{ATLAS-CONF-2015-029}. 
Finally, a track-based veto, the Jet Vertex Tagger (JVT), is used to reduce contributions from
jets arising from additional $pp$ interactions (pile-up)~\cite{ATLAS-CONF-2014-018}.

Jets including $b$-hadrons, referred as $b$-jets, are identified using 
information from a multivariate $b$-tagging algorithm that
combines information from an impact-parameter-based algorithm, from
the explicit reconstruction of an inclusive secondary vertex and from
a multi-vertex fitter that attempts to reconstruct the $b$- to
$c$-hadron decay chain~\cite{PERF-2012-04,b-tagging2016}. 
This $b$-tagging algorithm defines a set of ``$b$-tagged'' jets.
The working point used provides an efficiency to identify jets with
$b$-hadrons from the primary vertex of approximately $77\%$. The
rejection factors are 134 against light-quark and gluon jets, about 6
against jets originating from $c$-quarks, and about 22 against 
hadronically decaying $\tau$-leptons, as determined in a simulated
sample of top-quark pair ($\ttbar$) events~\cite{PERF-2012-04,b-tagging2016}. 
This $b$-tagging discriminant is used to categorize selected events as
discussed in section~\ref{sec:analysis}. The $b$-tagging algorithm
is also efficient in identifying jets containing $b$-hadrons that do
not originate from the primary vertex. The efficiency is largest for proper lifetimes
of $c\tau_a \sim\!0.5$~mm and decreases for longer lifetimes.

The missing transverse momentum, $\met$, is defined as the magnitude
of the transverse momentum imbalance
$\ensuremath{\vec{E}_{\text{T}}^{\text{miss}}}$, the negative vector
sum of the transverse momenta of calibrated selected objects, such as
electrons, muons and jets. The transverse momenta of charged-particle tracks compatible 
with the primary vertex and not matched to any of those objects are 
also included in $\ensuremath{\vec{E}_{\text{T}}^{\text{miss}}}$~\cite{Aaboud:2018tkc}.

Events are required to have at least one reconstructed electron or
muon with $\pt > 27$~\GeV\ which is matched within a cone of size
$\Delta R = 0.15$ to the lepton candidate reconstructed by the trigger
algorithms. Two event categories (single lepton and dilepton) are
used to probe $WH$ and $ZH$, respectively. 
Events with exactly one lepton are required to satisfy
$\met>25~{\GeV}$ and the transverse mass\footnote{The transverse
mass is defined as $\mtw \equiv \sqrt{2 \met \pt^{\ell} (1-
  \cos{\Delta \phi})}$, where $\pt^{\ell}$ is the transverse
momentum of the lepton and $\Delta\phi$ is the azimuthal angle between
the lepton and $\ensuremath{\vec{E}_{\text{T}}^{\text{miss}}}$ directions.} must
fulfil $\mtw >50~{\GeV}$, in order to be consistent with \Wboson\ boson
decays. Events in the dilepton channel must have exactly two
leptons with the same flavour and opposite electric charges. In the $ee$
and $\mu\mu$ channels, the dilepton invariant mass must be consistent
with the \Zboson\ boson mass window 85--100 \GeV. Events in the $e\mu$
channel (different flavour) are also used in the analysis to study backgrounds.  
Finally, events must have at least three jets, of which at least two
must be $b$-tagged. The selection requirements are summarized in 
table~\ref{tab:BaselineSelection}.

\begin{table}[ht!]
\centering
\caption{\label{tab:BaselineSelection} Summary of requirements for the
  single-lepton and dilepton channels. Here $m_{\ell\ell}$ is the
  dilepton invariant mass in the $ee$ and $\mu\mu$ channels.}
\begin{tabular}{lcc}
\toprule
Requirement & Single lepton & Dilepton \\
\midrule
Trigger & \multicolumn{2}{c}{single-lepton triggers}\\
Leptons & 1 isolated & 2 isolated, opposite-charge\\
Jets & \multicolumn{2}{c}{$\geq 3$} \\
$b$-tagged jets & \multicolumn{2}{c}{$\geq 2$} \\
Other & $\met>25~{\GeV}$, $\mtw >50~{\GeV}$ & $85~{\GeV} < m_{\ell\ell}  < 100~{\GeV}$\\
\bottomrule
\end{tabular}
\end{table}

\section{Signal and background modelling}
\label{sec:modelling}

Simulated event samples are used to study the
characteristics of the signal and to calculate its acceptance, as well
as for most aspects of the background estimation. 
Monte Carlo (MC) samples were produced using the full ATLAS
detector simulation~\cite{Aad:2010ah} based on {\GEANT4}~\cite{Agostinelli:2002hh}.
A faster simulation, where the full \GEANT4 simulation of the
calorimeter response is replaced by a detailed parameterization of the
shower shapes~\cite{ATL-PHYS-PUB-2010-013}, was adopted for some of
the samples. To simulate the effects of pile-up, additional interactions were
generated using \textsc{Pythia}~8.186~\cite{Sjostrand:2007gs} with
the A2 set of tuned parameters~\cite{Pythia8Tunes2012} and the
MSTW2008LO~\cite{Martin:2009iq} parton distribution function (PDF) set, and overlaid on
the simulated hard-scatter event. Simulated events were reweighted to
match the pile-up conditions observed in the data. All simulated events
are processed through the same reconstruction algorithms and analysis
chain as the data. In the simulation, the top-quark mass is assumed to
be $m_t=172.5$~\gev. Decays of $b$- and $c$-hadrons were performed by 
\textsc{EvtGen} v1.2.0~\cite{Lange:2001uf}, except in samples simulated
with the \textsc{Sherpa} event generator~\cite{Gleisberg:2008ta}.

Signal samples of associated Higgs boson production with a $\Wboson$
or $\Zboson$ boson, $pp \to WH$ or $ZH$, were generated with 
{\powheg}~v2 ~\cite{Nason:2004rx,Frixione:2007nw,Frixione:2007vw,Alioli:2010xd} 
using the CT10 PDF set~\cite{Lai:2010vv}
at next-to-leading order (NLO). The Higgs boson mass is assumed to be $m_H =
125~{\GeV}$. 
The Higgs boson decay into two spin-zero $a$-bosons and the subsequent
decay of each $a$-boson into a pair of
$b$-quarks were simulated with \pythia~8.186.
The $a$-boson decay was performed in the narrow-width
approximation and the coupling to the $b$-quarks is assumed to be that of
a pseudoscalar. However, since the polarization of the quarks is not
observable, this search is insensitive to the chosen parity
hypothesis for the $a$-boson. \pythia~8.186 was also used for the showering,
hadronization, and underlying-event (UE) simulation with the A14
tune~\cite{Pythia8Tunes2014}. 
The mass of the $a$-boson was varied for different signal hypotheses in the range $20~{\GeV} \leq
m_a \leq 60~{\GeV}$, in $10~{\GeV}$ mass steps. 

Signal samples with long-lived $a$-bosons were generated with
\amcatnlo~\cite{Alwall:2014hca} at leading order (LO) using the
NNPDF2.3LO~\cite{Ball:2014uwa} PDF set and showered with
\pythia~8.186. The model used is the SM with an additional dark
sector that includes a dark vector boson and a dark
Higgs boson~\cite{Curtin:2014cca,Curtin:2013fra,deFlorian:2016spz}.
In this model, the dark Higgs boson, which plays the role of the
$a$-boson, is a scalar under parity transformation and decays promptly.  
Therefore, the lifetimes of the $a$-bosons were replaced
with values sampled randomly from an exponentially falling
distribution with the desired mean value.  Signal MC samples were
produced for $a$-boson mean proper lifetimes of 0.1, 1, and 10~mm.
Intermediate $a$-boson lifetimes can be obtained by reweighting these
samples.  The masses of the $a$-boson are 20, 30, and 60~GeV.  The
uneven spacing of $a$-boson masses is motivated by the fact that the
signal kinematics (and therefore acceptance) change significantly
between 20 and 30~GeV, but are quite similar from 30 to 60~GeV.  The
ATLAS fast detector simulation was used for samples of long-lived
$a$-bosons, after verifying that it correctly reproduces the \GEANT4-based
simulation for the range of $a$-boson lifetimes under consideration.

The sample used to model the \ttbar\ background was generated using the
{\powheg}~v2 event generator~\cite{Powhegttbar}, with the NNPDF3.0NLO PDF
set. The {\powheg} model parameter $h_{\mathrm{damp}}$, which controls matrix
element (ME) to parton shower (PS) matching and effectively
regulates the high-$\pt$ radiation, was set to $h_{\mathrm{damp}} = 1.5 m_t $~\cite{Top2016}. 
The PS and the hadronization were modelled by  
\pythia~8.210~\cite{Sjostrand:2014zea} with the A14 tune. 
The renormalization and factorization scales were set to the transverse mass of the top quark, 
defined as $m_{\textrm{T},t} = \sqrt{m_t^2 + p_{\textrm{T},t}^2}$, where $p_{\textrm{T},t}$ 
is the transverse momentum of the top quark in the \ttbar\ centre-of-mass reference frame.
The $\ttbar$ sample is normalized to the next-to-next-to-leading-order
(NNLO) theoretical cross-section of $832^{+46}_{-51}$~pb, obtained
with \textsc{Top++}~2.0~\cite{Czakon:2011xx}. 
Alternative \ttbar\ samples used to derive systematic uncertainties
are described in section~\ref{sec:systematics}.

The simulated $\ttbar$ events are categorized depending on the
parton-level flavour content of particle jets\footnote{Particle jets are reconstructed by
 clustering stable particles, excluding muons and neutrinos, using the
 anti-$k_t$ algorithm with a radius parameter $R=0.4$.} not
originating from the decay of the $\ttbar$ system, using the 
procedure described in refs.~\cite{Aad:2015gra,Aaboud:2017rss}. Events containing
at least one particle jet matched to a $b$-hadron are labelled
as $\ttbb$. Events containing at least one particle jet
matched to a $c$-hadron and no $b$-hadron are labelled as $\ttcc$. The 
$\ttbb$ and $\ttcc$ categories are generically referred to as 
$\ttbar$+HF events (with HF standing for ``heavy flavour''). Remaining 
events are labelled as $\ttbar$ + light-jets (referred to as $\ttlight$)
and also include events with no additional particle jets. 

To model the dominant \ttbb\ background with the highest available precision, 
the relative contributions of the different heavy-flavour categories
in the $\ttbar$ sample described above
are scaled to match the predictions of an NLO \ttbb\ 
sample including parton showering and
hadronization~\cite{Cascioli:2013era}, generated with 
\textsc{Sherpa+OpenLoops}~\cite{Gleisberg:2008ta, Cascioli:2011va},
using the procedure described in ref~\cite{Aad:2015gra}.
The sample was produced with \SHERPA~2.2.1 and the CT10
four-flavour (4F) scheme PDF set~\cite{Re:2010bp,Alioli:2009je}.
The renormalization scale for this sample was chosen to be
$\mu_{\mathrm{R}} = \prod_{i=t,\bar{t},b,\bar{b}}
E_{\mathrm{T},i}^{1/4}$ using the CMMPS
prescription~\cite{Cascioli:2013era}, while the factorization
scale is set to $\mu_{\mathrm{F}}  = \frac{1}{2}
\sum_{i=t,\bar{t},b,\bar{b}} E_{\mathrm{T},i}$. The resummation scale
$\mu_{\mathrm{Q}}$, which sets an upper bound for
the hardness of the PS emissions, was also set to $\frac{1}{2}
\sum_{i=t,\bar{t},b,\bar{b}} E_{\mathrm{T},i}$. 

The production of \Wboson and \Zboson bosons in association with jets was
simulated with \SHERPA~2.2.1~\cite{Gleisberg:2008ta} using the
NNPDF3.0NNLO PDF set for both the ME 
calculation and the dedicated PS tuning developed
by the \SHERPA authors~\cite{Schumann:2007mg}.  The ME
calculation was performed with \textsc{Comix}~\cite{Gleisberg:2008fv} and
\textsc{OpenLoops}~\cite{Cascioli:2011va}, and was matched to the \SHERPA
PS using the MEPS@NLO prescription~\cite{Hoeche:2012yf}.
The MEs were calculated for up to two additional partons at
NLO and for three and four partons at LO in QCD.
The $W/Z$~+~jets samples are normalized to the NNLO
cross-sections~\cite{Melnikov:2006kv, ATLAS-CONF-2015-039}. 

The diboson~+~jets samples were generated using \textsc{Sherpa} 2.1.1 
as described in ref.~\cite{multibosonMC}.
Samples of $\ttbar \Wboson$ and $\ttbar \Zboson$ ($\ttbar V$) events 
were generated with an NLO ME using {\amcatnlo} interfaced to
\pythia~8.210 with the NNPDF3.0NLO PDF set and the A14 tune.

Samples of $Wt$ and $s$-channel single-top-quark backgrounds were generated with 
{\powheg}~v1 at NLO accuracy using the CT10 PDF set.
Overlap between the \ttbar\ and $Wt$ final states was resolved using
the ``diagram removal'' scheme~\cite{Frixione:2008yi}.
The $t$-channel single-top-quark events 
were generated using the {\powheg}~v1 event generator at NLO accuracy with 
the CT10 4F scheme PDF set. For this process, top quarks were 
decayed using \textsc{MadSpin}. All single-top-quark samples were
interfaced to \pythia~6.428~\cite{Sjostrand:2006za} with the
Perugia 2012 tune~\cite{Skands:2010ak}.
The single-top quark $t$- and $s$-channel samples are normalized to
the NLO theoretical cross-sections~\cite{ttbarxs,Kant:2014oha}, while the $Wt$ channel is normalized to
the approximate NNLO prediction~\cite{Kidonakis:2010ux,Kidonakis:2013zqa}.

Higgs boson production in association with a single top quark is a
rare process in the SM, but is included in the analysis and is treated as a background. 
Samples of single top quarks produced in association with a $W$ boson
and a Higgs boson, $tWH$, were produced with \amcatnlo\ interfaced to
\herwigpp~\cite{Herwigpp} with the CTEQ6L1 PDF set. The other Higgs
boson production modes are found to be negligible and are not considered.
The production of four top quarks ($t\bar{t}t\bar{t}$) as well as $t\bar{t}WW$ events
were generated with \amcatnlo{} with LO accuracy and interfaced to \pythia~8.186. 
The $tZW$ production process was also generated with \amcatnlo\ interfaced to
\pythia~8.186, but at NLO accuracy.

In the single-lepton channel, the background from events with a jet or
a photon misidentified as a lepton or with non-prompt leptons from
hadron decays (hereafter referred to as a fake or non-prompt lepton)
is estimated directly from data using a matrix method~\cite{ATLAS-CONF-2014-058}. 
A data sample enhanced in fake and non-prompt leptons is selected by
removing the lepton isolation requirements and, for electrons,
loosening the identification criteria. The efficiency for these
``loose'' leptons to satisfy the nominal selection (``tight'') criteria is
measured in data, separately for real prompt leptons and for fake or
non-prompt leptons. For real prompt leptons, the efficiency is
measured in $Z$ boson events, 
while for fake and non-prompt leptons, it is estimated from events
with low $\met$ and low values of $\mtw$.

The number of fake or non-prompt leptons satisfying the tight criteria
can then be calculated by inverting the matrix defined by the two equations:
\[
N^{\textrm{l}} = N_{\textrm{r}}^{\textrm{l}} + N_{\textrm{f}}^{\textrm{l}},
\qquad
N^{\textrm{t}} = \varepsilon_{\textrm{r}} N_{\textrm{r}}^{\textrm{l}} + \varepsilon_{\textrm{f}} N_{\textrm{f}}^{\textrm{l}},
\]
where $N^{\textrm{l}}$ ($N^{\textrm{t}}$) is the number of events
in data satisfying the loose (tight) lepton selection, 
$N_{\textrm{r}}^{\textrm{l}}$ ($N_{\textrm{f}}^{\textrm{l}}$) is the
number of events with a real prompt (fake or non-prompt) lepton in the loose lepton sample, 
and $\varepsilon_{\textrm{r}}$ ($\varepsilon_{\textrm{f}}$) is the
efficiency for these events to fulfil the tight lepton selection.
By generalizing the resulting formula to extract $\varepsilon_{\textrm{f}} N_{\textrm{f}}^{\textrm{l}}$, 
a weight is assigned to each event selected in the loose lepton data sample, 
providing a prediction for both the yields and the kinematic
distributions of the fake and non-prompt lepton background. 

When applying the matrix method in the case of high jet and $b$-tagged
jet multiplicities, the number of events in data satisfying the loose and tight
lepton selections is significantly reduced, leading to large
fluctuations in the background predictions. In order to mitigate this
problem, instead of tagging the jets by applying the $b$-tagging
algorithm, their probabilities to be $b$-tagged are parameterized as a
function of the jet $\pt$. This allows all events in the sample before
$b$-tagging is applied to be used in predicting the normalization and
shape of the background from fake or non-prompt leptons after
$b$-tagging. The tagging probabilities are derived using
an inclusive sample of fake or non-prompt leptons and the resulting
predictions of this background estimate are in agreement with those
obtained by applying the $b$-tagging algorithm and have greatly reduced
statistical uncertainties.

In the dilepton channel, the background contribution from
fake or non-prompt leptons is very small and is estimated from simulation
and normalized to data in a control region with two same-charge leptons.

\section{Event categorization}
\label{sec:analysis}

Events satisfying the object selection are categorized into analysis regions
according to the number of leptons, jets and $b$-tagged jets. The
regions enhanced in signal $H \to aa \to 4b$ events relative to
the backgrounds are referred to as signal regions (SRs). The other regions,
referred to as control regions (CRs), are used to constrain the background
predictions and related systematic uncertainties (see
section~\ref{sec:systematics}) through a profile likelihood 
fit to the data (see section~\ref{sec:results}). The signal and
backgrounds are derived consistently in the signal and control regions
in a combined fit. The discrimination of signal from background is
further enhanced in the signal regions by using multivariate
techniques, as described in section~\ref{sec:analysisStrategy}.  

\begin{figure*}[htbp]
\begin{center}
        \includegraphics[width=\textwidth]{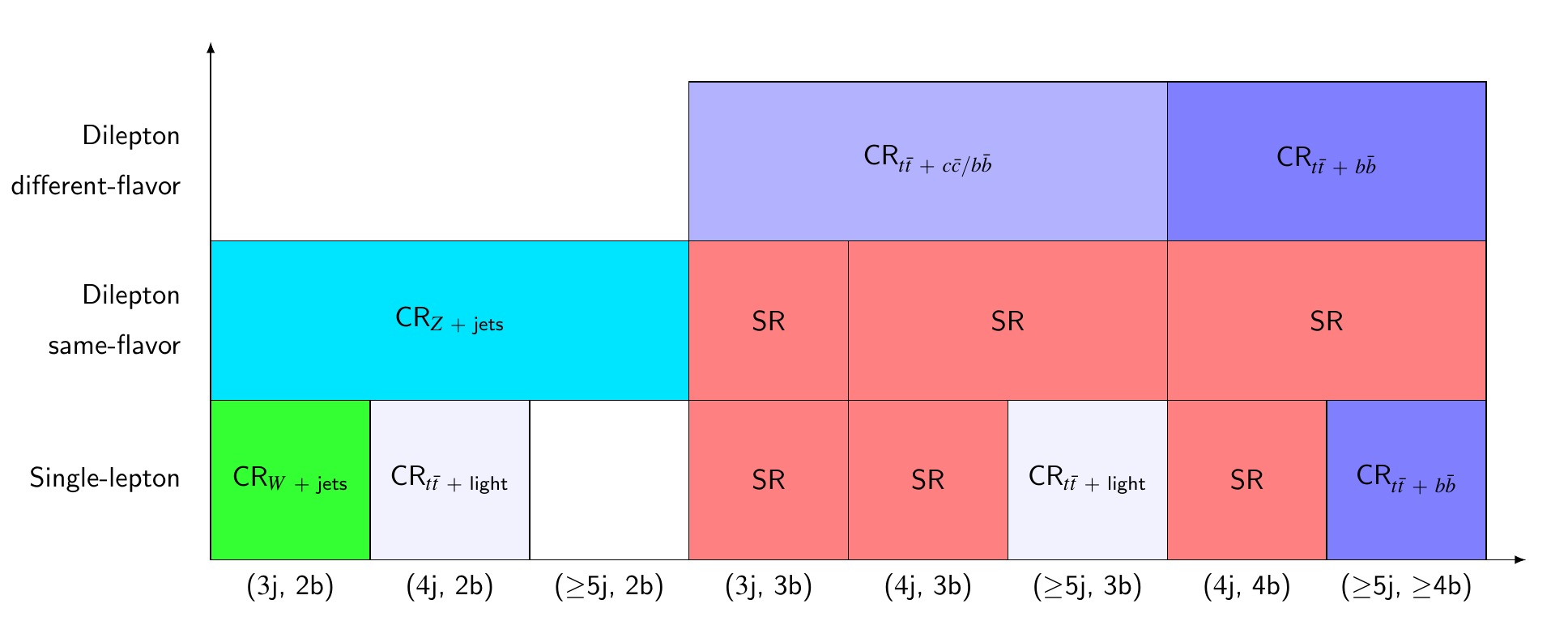}\\
        \caption{Definition of the signal and control regions (SR and
          CR, respectively) in the
          single-lepton and dilepton channels. The main background
          component probed with the CR is indicated. The vertical axis
          shows the lepton selection, while the horizontal axis shows
          the jet and $b$-tagged jet multiplicities.}
    \label{fig:regions}
\end{center}
\end{figure*}

\begin{figure*}[!ht]
\subfloat[]{\includegraphics[width=0.47\textwidth]{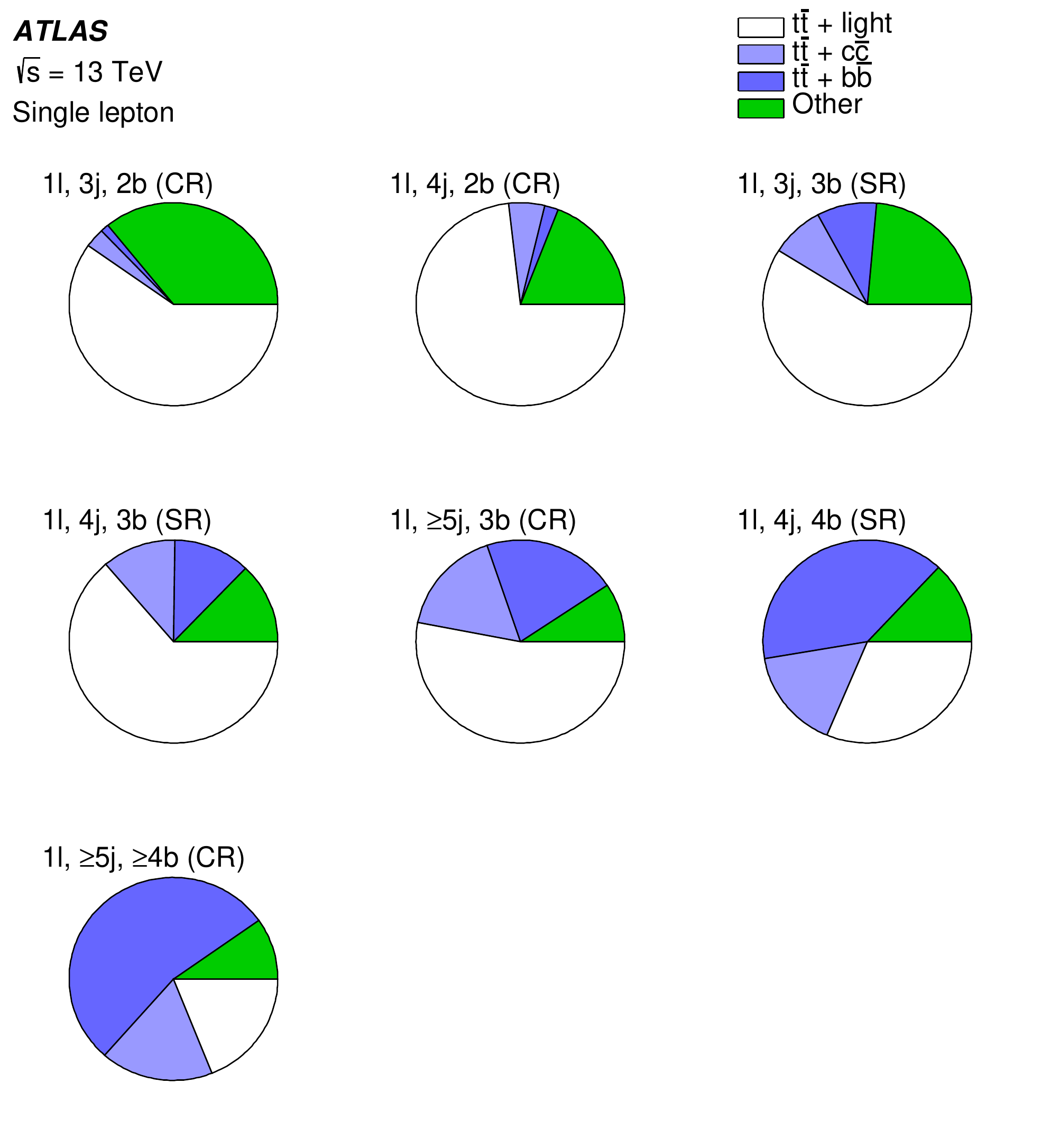}}%
\hfill
\subfloat[]{\raisebox{15.5ex}{\includegraphics[width=0.47\textwidth]{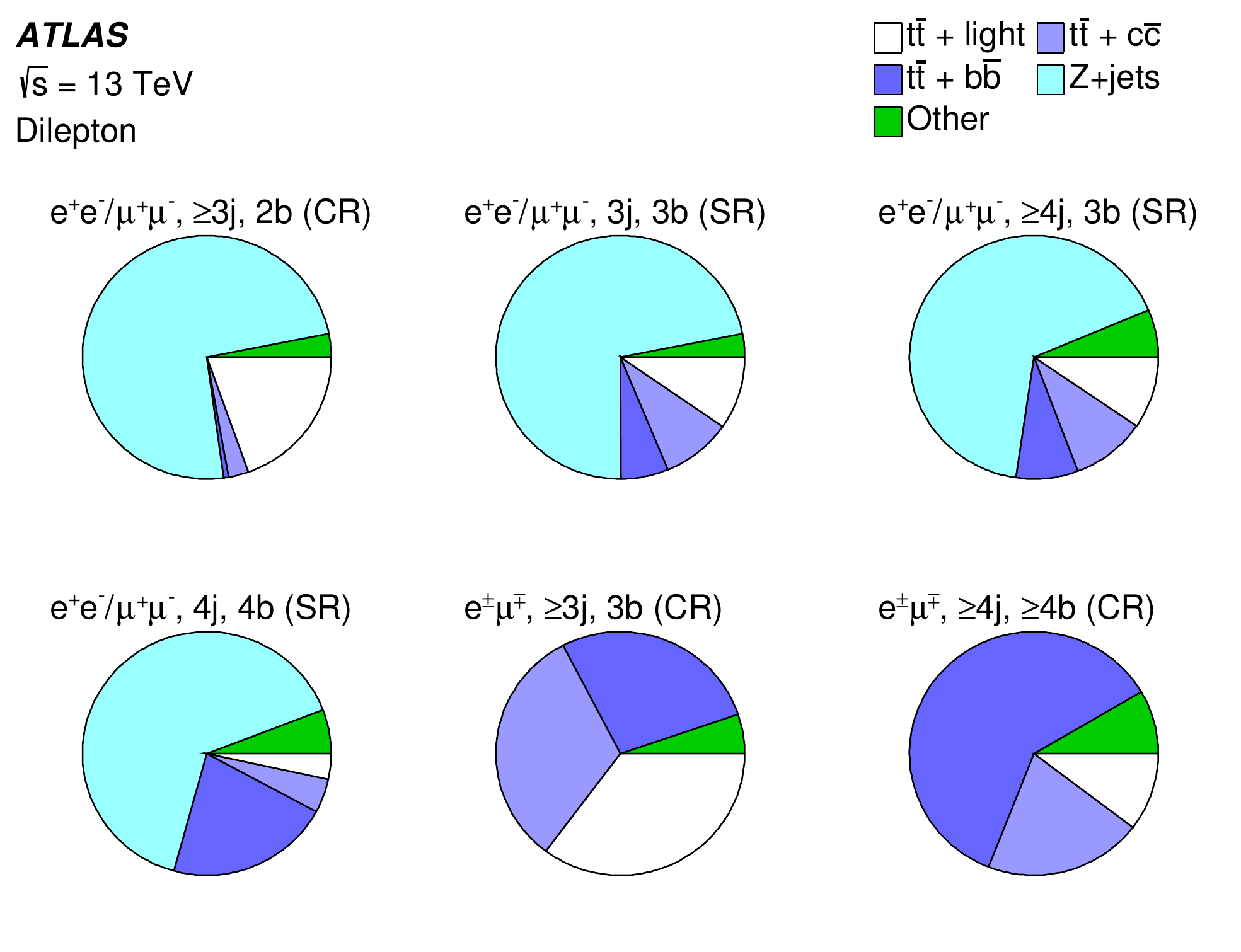}}}\\
\caption{Fractional contributions of the various backgrounds to the
  total background prediction in the (a) single-lepton and (b)
  dilepton signal and control regions.  The predictions for the
  various background contributions are obtained through the simulation 
  and the data-driven estimates described in
  section~\ref{sec:modelling}. The $\ttbar$ background categories are
  also defined in section~\ref{sec:modelling}.}
\label{fig:piechart}
\end{figure*}

The $H \to aa \to 4b$ decay chain is expected to have multiple
$b$-tagged jets, often three or four, satisfying the object
selection. The dominant background arises from \ttbar\ events in the 
single-lepton channel and  $Z$~+~jets events in the dilepton channel, which can
also have different jet and $b$-tagged jet multiplicities or leptons
of different flavour in the case of the dilepton channel. 
The regions are referred to as ($n_{\ell} \ell$, $n_j$j, $n_b$b)
indicating $n_{\ell}$ leptons, $n_j$ selected jets and $n_b$ $b$-tagged jets.
The SRs contain at least three $b$-tagged jets and are (1$\ell$, 3j,
3b), (1$\ell$,4j, 3b) and (1$\ell$, 4j, 4b) for single-lepton events,
and (2$\ell$, 3j, 3b), (2$\ell$, $\geq$4j, 3b) and (2$\ell$, $\geq$4j,
$\geq$4b) for same-flavour dilepton events. 
The CRs are (1$\ell$, 3j, 2b), (1$\ell$, 4j, 2b), (1$\ell$, $\geq$5j,
3b) and (1$\ell$, $\geq$5j, $\geq$4b) for single-lepton events,
(2$\ell$, $\geq$3j, 2b) for same-flavour dilepton events, and
(2$\ell$, $\geq$3j, 3b) and (2$\ell$, $\geq$4j, $\geq$4b) for
different-flavour dilepton events. The signal and control regions are
summarized in figure~\ref{fig:regions}, indicating the main background 
sources probed in the CRs. Figure~\ref{fig:piechart} summarizes the
background composition in the signal and control regions.

In the single-lepton signal regions, background $\ttbar$ events can 
only satisfy the selection criteria if accompanied by additional 
$b$-tagged jets. 
The $\ttlight$ background is dominant 
in the sample of events with exactly two or three $b$-tagged jets. The background
processes $\ttcc$ and $\ttbb$ become more important as the jet and
$b$-tagged-jet multiplicities increase. In particular, the $\ttbb$ background
dominates for events with $\geq$5 jets and $\geq$4 $b$-tagged jets. 
In the case of (1$\ell$, 3j, 3b) or (1$\ell$, 4j, 3b), the main sources of
$\ttbar$ background are events with jets mistagged as $b$-tagged jets,
particularly from $W \to cs$ decays, where the $c$-jet is
misidentified, and from $W \to \tau\nu$, where the
$\tau$-lepton decays hadronically and is likewise mistagged. 
In the case of (1$\ell$, 4j, 4b), the $\ttbar$ background includes
more events with genuine $b$-quarks from gluon splitting to $\bbbar$ pairs.

In the dilepton channel, the background is mainly composed of $\ttbar$
and \Zjets{} events. In the case of the $\ttbar$ background, most
events contain two prompt leptons from the leptonic decays of the two $W$
bosons, and two $b$-jets from the top-quark decays. Additional jets
arise from gluon splitting into $\bbbar$ and $\ccbar$ and from
jets from initial-state radiation and pile-up. In each of
these cases, the third and fourth $b$-tags in the event are from
additional $b$-tagged jets, or from the mistag of additional $c$- or
light-jets. In the samples with exactly three or four jets and exactly
three $b$-tagged jets, the contributions
of each of these sources is similar. In the case of the sample with
exactly four jets and exactly four $b$-tagged jets, the contribution from events
with real additional $b$-tagged jets, such as from gluon splitting into
$\bbbar$, dominates. 

In the case of the \Zjets{} background, the dominant contribution is
from $Z$ bosons produced in association with multiple heavy-flavour 
jets, for example from gluon splitting into $\bbbar$. In particular,
for events with exactly three jets and three $b$-tagged jets or
exactly four jets and four $b$-tagged jets, about half of the events
are from $Z+\bbbar$ with a mistagged light-flavoured jet and half are
from $Z+\bbbar c$ with a mistagged $c$-jet. In the case of the events with 
exactly four jets and three $b$-tagged jets, approximately a third of
the events are from $Z+\bbbar$ with a mistagged light-flavoured jet, a third
are from $Z+\bbbar c$ with a mistagged $c$-jet and a third are from
$Z+\bbbar b$. 

In the dilepton channel, the control regions are designed to be
populated by the two main background processes: \ttbar{} and
\Zjets{}. The control region with two
same-flavour leptons, (2$\ell$, $\geq$3j, 2b), is populated by \Zjets{}
and $\ttlight$. The control regions with two different-flavour
leptons but with the same jet and $b$-tagged jet multiplicities as
the signal regions, (2$\ell$, $\geq$3j, 3b) and (2$\ell$, $\geq$4j,
$\geq$4b), are enhanced in $\ttcc$ and $\ttbb$ processes.

\section{Analysis strategy}
\label{sec:analysisStrategy}

In each of the six signal regions, a boosted decision tree (BDT)
discriminant that combines information from several 
variables provides additional discrimination between signal and background.
In the control regions, kinematic variables are used to
provide additional discrimination between distinct sources of background.
The distribution of $\HT$, defined as
the scalar sum of the $\pt$ of the jets, is used in the control
regions with two~$b$-tagged jets, while the invariant mass of the $b$-tagged
jets is used in the control regions with three or four $b$-tagged jets.
A statistical analysis based on a binned likelihood function
constructed as a product of Poisson probability terms over all regions
and all bins considered in the search is used to derive the background
predictions and uncertainties, and to test for the presence of signal.
The statistical procedure and the results are described in section~\ref{sec:results}.

The toolkit for multivariate analysis (TMVA)~\cite{Hoecker:TMVA} 
is used to train the BDT discriminant. In the single-lepton channel,
dedicated BDTs are trained to discriminate each of the signals with
$a$-boson masses of 20, 30 and 50~\GeV\ from $\ttbar$
events. The discriminant trained at 50~\GeV\ is found to have good
sensitivity for the $a$-boson mass range 40--60~\GeV.  In the dilepton
channel, dedicated BDTs are trained to discriminate each of the
signals from both the $\ttbar$ and \Zjets{} events. The discriminant
trained at 30~\GeV\ is found to have good sensitivity over the full
mass range (20--60~\GeV) for each of the signal regions. 

The choice of inputs used in the BDT discriminants was based on the following considerations.
Signal events are characterized by the
presence of a resonance resulting from the Higgs boson decay $H \to aa
\to 4b$. Several variables are used to reconstruct the particles from the signal
decay chain. The first is the reconstructed invariant mass of the
$b$-tagged jets, {$m_{bbb}$} or {$m_{bbbb}$}, defined for events with
three or four $b$-tagged jets respectively, which peaks around the
Higgs boson mass for signal events. In the case of three $b$-tagged
jets, the peak is due to events where two $b$-quarks are merged in a
single jet or one of the $b$-quarks is very soft in an asymmetric
decay and has a small impact on the kinematics. In the case of events
with four $b$-tagged jets, the invariant masses of the two $b$-tagged
jet pairs are discriminating variables between signal and
background. The pairings are chosen to minimize the difference between
the invariant masses of the $b$-tagged jet pairs, and are labelled
$m_{bb1}$ and $m_{bb2}$, such that $m_{bb1} > m_{bb2}$.

Additional kinematic variables exhibit differences between signal and
background. In both channels, the average angular separation between 
$b$-tagged jets, referred to as average $\Delta R$($b$,$b$), is typically larger for
background events where the $b$-tagged jets originate from the decays
of different particles, such as the two top quarks in $\ttbar$
events. In the single-lepton channel, several additional kinematic variables
are included in the BDT discriminant. The $\HT$ variable is a
measure of the total hadronic energy in the event, which is typically
larger for $\ttbar$ than for $WH$ events. 
The transverse momentum of the $W$ boson, $\pt^W$, 
constructed from the vector sum of the
$\ensuremath{\vec{E}_{\text{T}}^{\text{miss}}}$ and the
lepton $\ensuremath{\vec{p}_{\text{T}}}\xspace$, is slightly higher
for signal $WH$ events, where the $W$ boson recoils against the Higgs
boson, than for background \ttbar events.

Finally, two other variables are used to identify particles from the dominant $\ttbar$
background decay chain in the single-lepton channel. The first
variable is used in the (1$\ell$, 4j, 3b) channel to select
$\ttbar$ events with two $b$-tagged jets from the top-quark decays and
a third $b$-tagged jet from a misidentified $c$- or light-jet from the
hadronically decaying $\Wboson$ boson. This variable is the invariant
mass of two $b$-tagged jets (selected as the pair with the smallest
$\Delta R$ separation) and the non-$b$-tagged jet, $m_{bbj}$, which
reconstructs the hadronically decaying top quark, peaking
around the top-quark mass for these background events. 
The second variable is an  $m_{\mathrm{T2}}$ observable, defined as
the minimum ``mother'' particle mass compatible 
with all transverse momenta and mass-shell
constraints~\cite{Burns:2008va}, that identifies events with several
invisible particles. In the case of the \ttbar background events, in
addition to the \met from the neutrino from a leptonic $W$ boson
decay, invisible particles may arise from a $\tau$-lepton decay or from a
lost jet from a $W$ boson. In these cases, $m_{\mathrm{T2}}$ has an
endpoint at the top-quark mass, which is not the case for the signal.  

For the dilepton channel, two variables are sensitive to the
signal topology of a $Z$ boson recoiling against a Higgs boson: 
the separation between the two leptons in the event, $\Delta
R$($\ell$,$\ell$), and the separation between the $Z$ boson,
constructed from the two leptons, and the Higgs boson, constructed
from the $b$-tagged jets, $\Delta R$($\Zboson$,$H$). Another
discriminating variable that carries information about the signal production
mechanism is the cosine of the polar angle of the Higgs
boson in the reference frame of the parent process $Z^* \rightarrow
ZH$, referred to as $\cos{\theta^*}$, which is sensitive to the spin of the parent
particle. Finally, the $\met$ variable is used to discriminate against
background $\ttbar$ events that include the presence of multiple 
neutrinos. 

\begin{table}[h!]
\centering
\caption{\label{tab:BDTVars} List of variables used to train the BDT
  multivariate discriminant for each signal region.}
  \begin{tabular*}{\textwidth}{@{\extracolsep{\fill}}l|ccc|ccc}
\toprule
Variable & (1$\ell$, 3j, 3b)  & (1$\ell$, 4j, 3b) & (1$\ell$, 4j, 4b) & (2$\ell$, 3j, 3b) & (2$\ell$, $\geq$4j, 3b) & (2$\ell$, $\geq$4j, $\geq$4b)\\
\midrule
$m_{bbb}$ & \checkmark & \checkmark & & \checkmark & \checkmark & \\
$m_{bbbb}$ & & & \checkmark & & & \checkmark \\
$m_{bb1}$ & & & \checkmark & & & \checkmark\\
$m_{bb2}$ & & & \checkmark & & & \checkmark\\
Average $\Delta R$($b$,$b$) & \checkmark & \checkmark & \checkmark & \checkmark & \checkmark & \checkmark \\
$\HT$ & \checkmark & \checkmark & \checkmark & & & \\
$\pt^W$ & \checkmark & & & & & \\
$m_{bbj}$ & & \checkmark & & & & \\
$m_{\mathrm{T2}}$ & \checkmark & \checkmark & \checkmark & & & \\
$\Delta R$($\ell$,$\ell$) & & & & \checkmark & \checkmark & \checkmark \\
$\Delta R$($\Zboson$,$H$) & & & & \checkmark & \checkmark & \\
$\cos{\theta^*}$ & & & & & & \checkmark \\
$\met$ & & & & \checkmark & \checkmark & \checkmark \\
\bottomrule
\end{tabular*}
\end{table}

\begin{figure*}[htbp]
\begin{center}
        \subfloat[]{\includegraphics[width=0.33\textwidth]{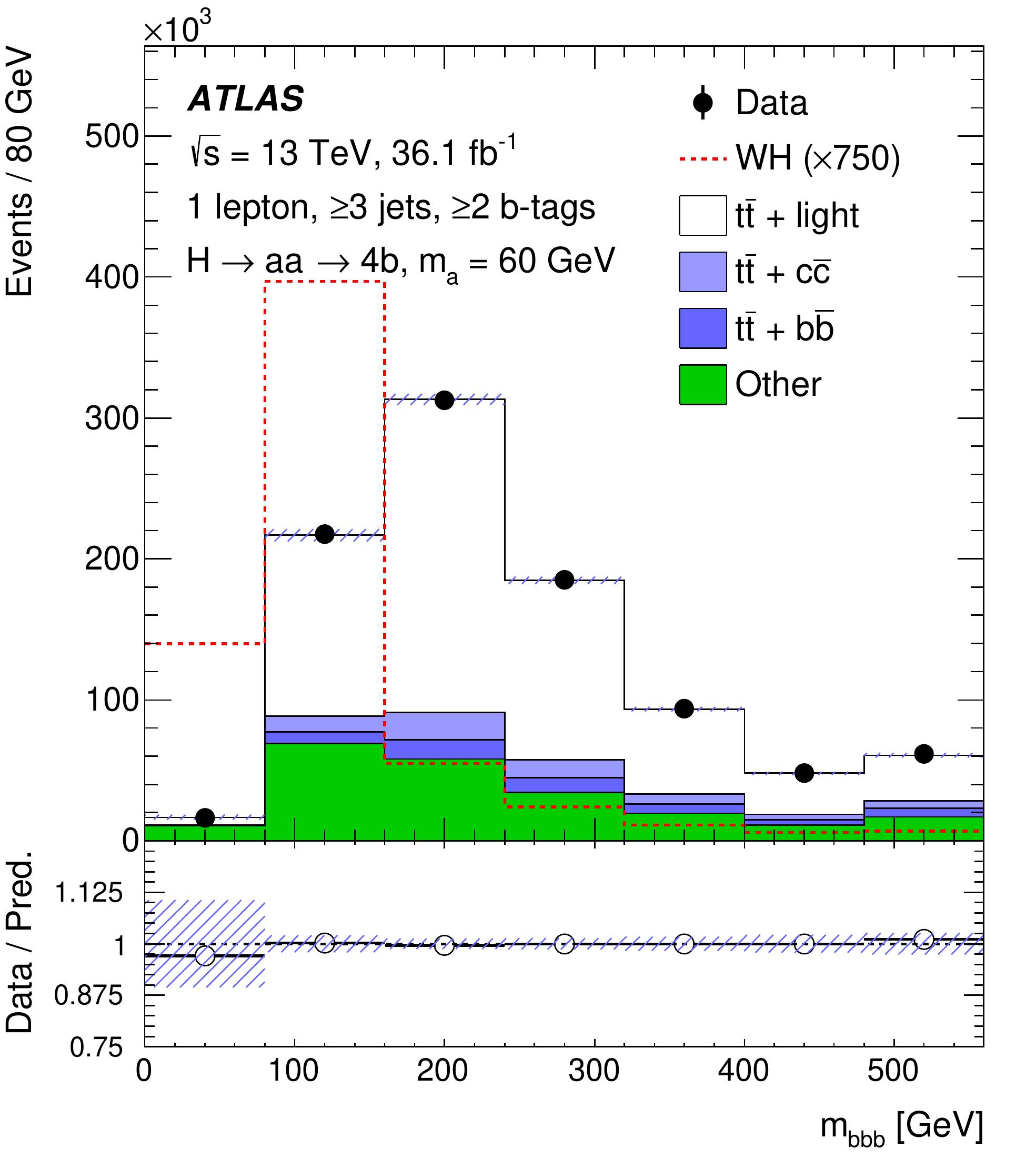}}%
        \subfloat[]{\includegraphics[width=0.33\textwidth]{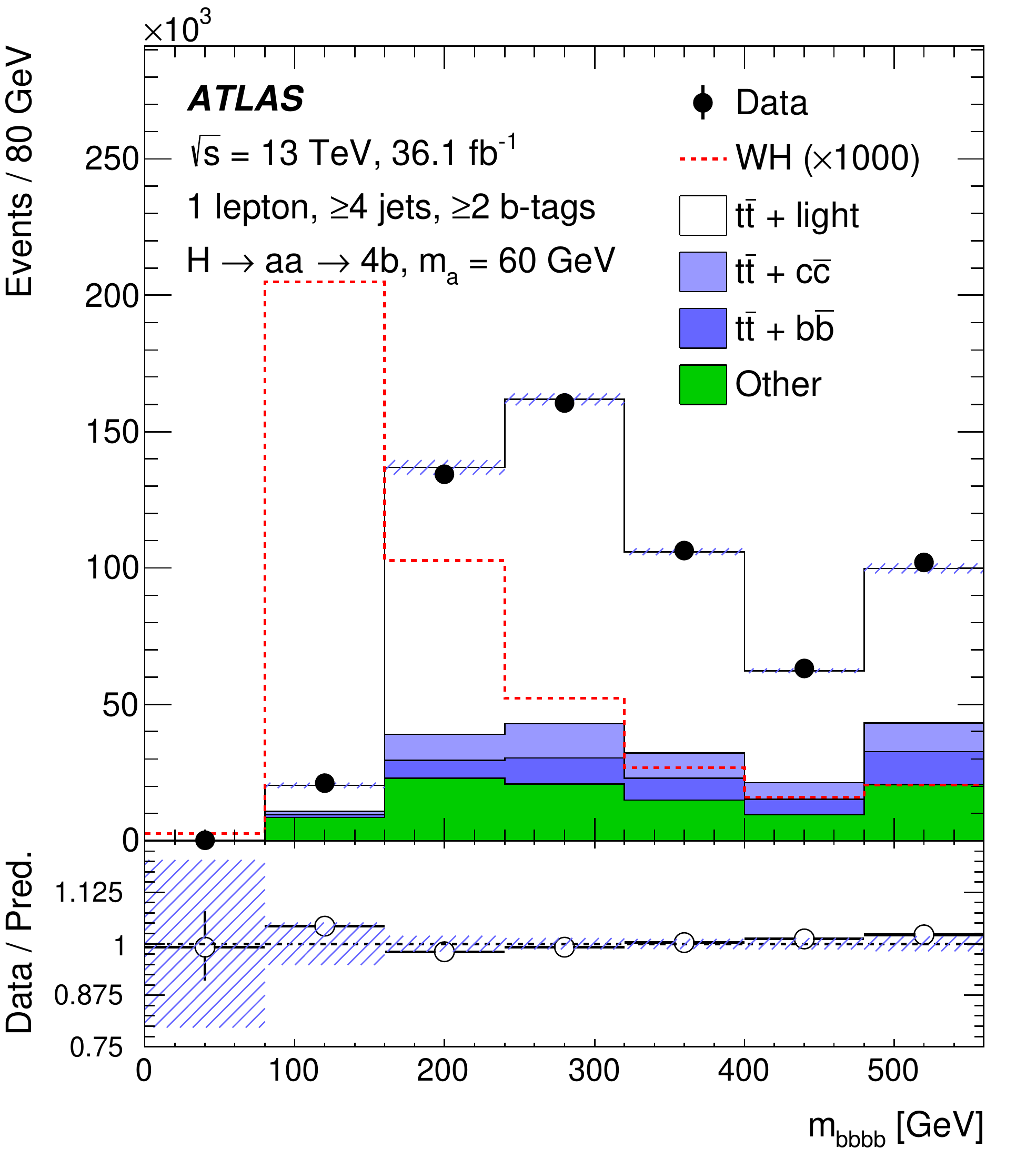}}\\
        \subfloat[]{\includegraphics[width=0.33\textwidth]{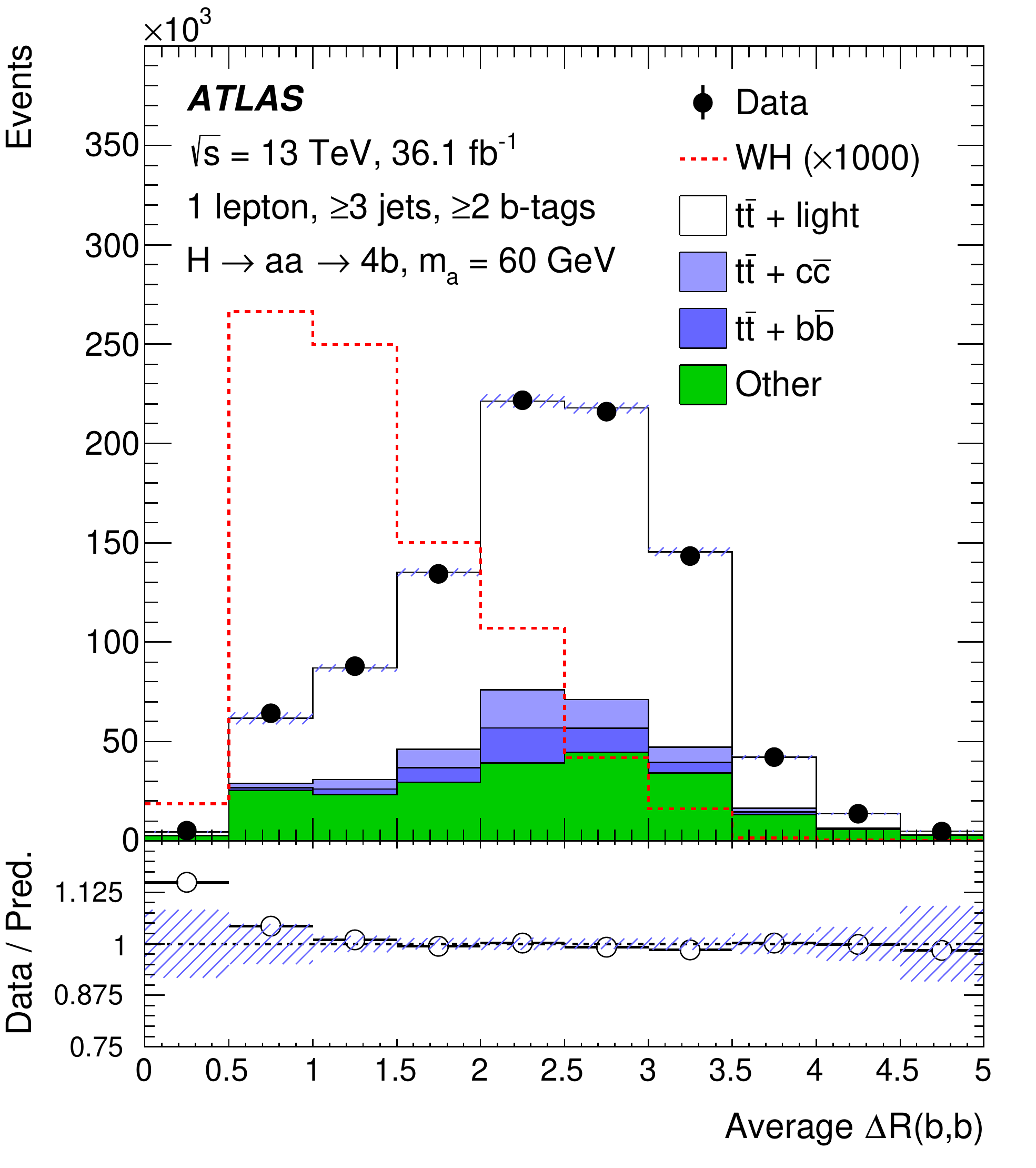}}%
        \subfloat[]{\includegraphics[width=0.33\textwidth]{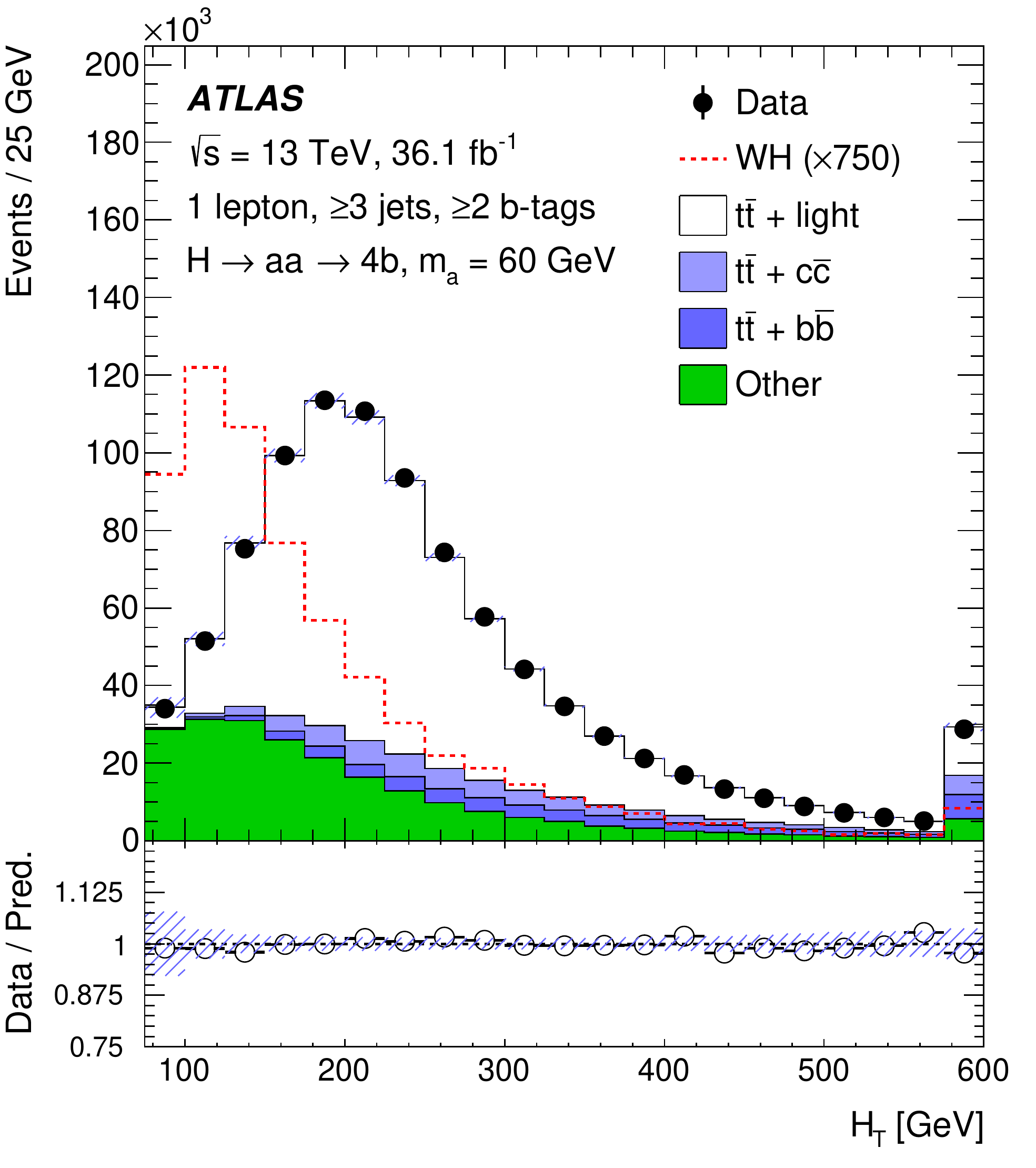}}\\
        \subfloat[]{\includegraphics[width=0.33\textwidth]{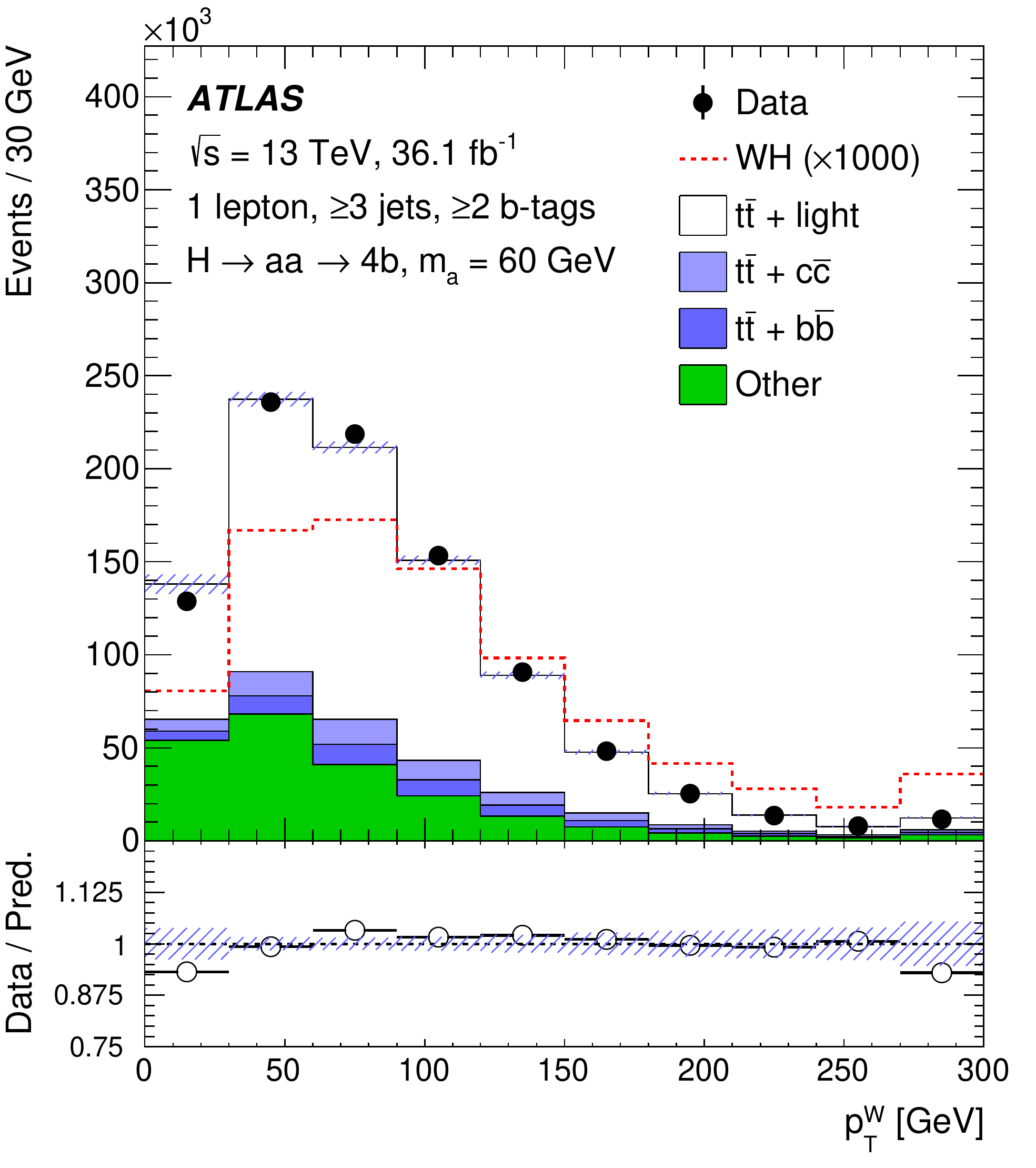}}%
        \subfloat[]{\includegraphics[width=0.33\textwidth]{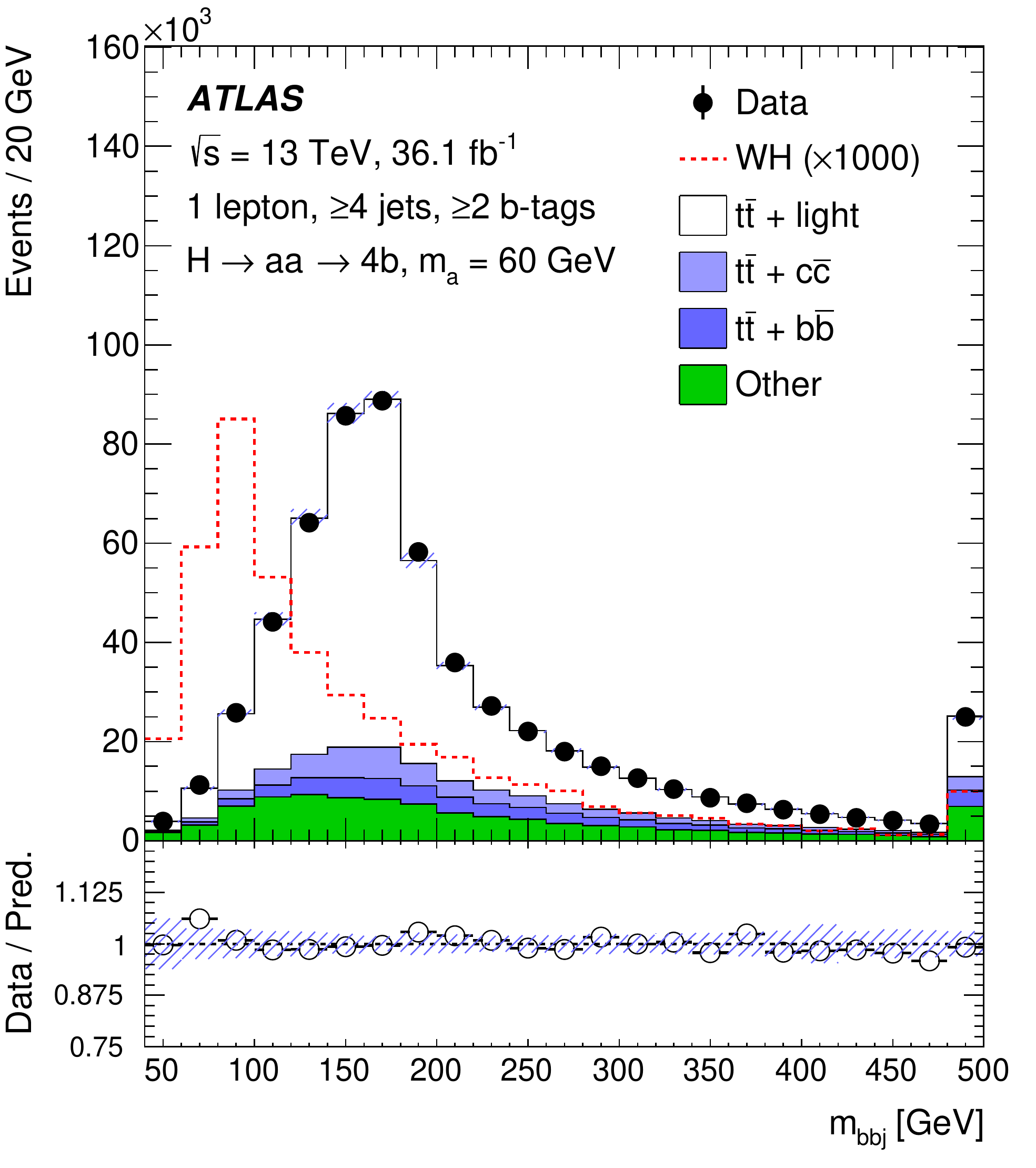}}\\
        \caption{Comparison of data with the post-fit background
          estimates for (a) $m_{bbb}$, (b)
          $m_{bbbb}$, (c) average $\Delta R$($b$,$b$),
          (d) $\HT$, (e) $\pt^W$ and
          (f) $m_{bbj}$ in the single-lepton sample inclusive in
          number of jets and $b$-tagged jets. Comparisons use events with 
          $\ge3$ jets, except when $\ge4$ jets are necessary to define
          the variable, in which case events with $\ge4$ jets are used. Distributions for the signal model
          ($WH$, $H\rightarrow aa \rightarrow 4b$), with $m_a=$ 60
          GeV, normalized to the SM $pp \rightarrow WH$ cross-section, assuming $\BR (H\rightarrow aa
          \rightarrow 4b) = 1$ and scaled by a factor as indicated in
          the figure, are overlaid. The hashed area represents the total uncertainty in the
          background. The last bin contains the overflow.}
    \label{fig:kinvar_validation_1}
\end{center}
\end{figure*}

\begin{figure*}[htbp]
\begin{center}
        \subfloat[]{\includegraphics[width=0.33\textwidth]{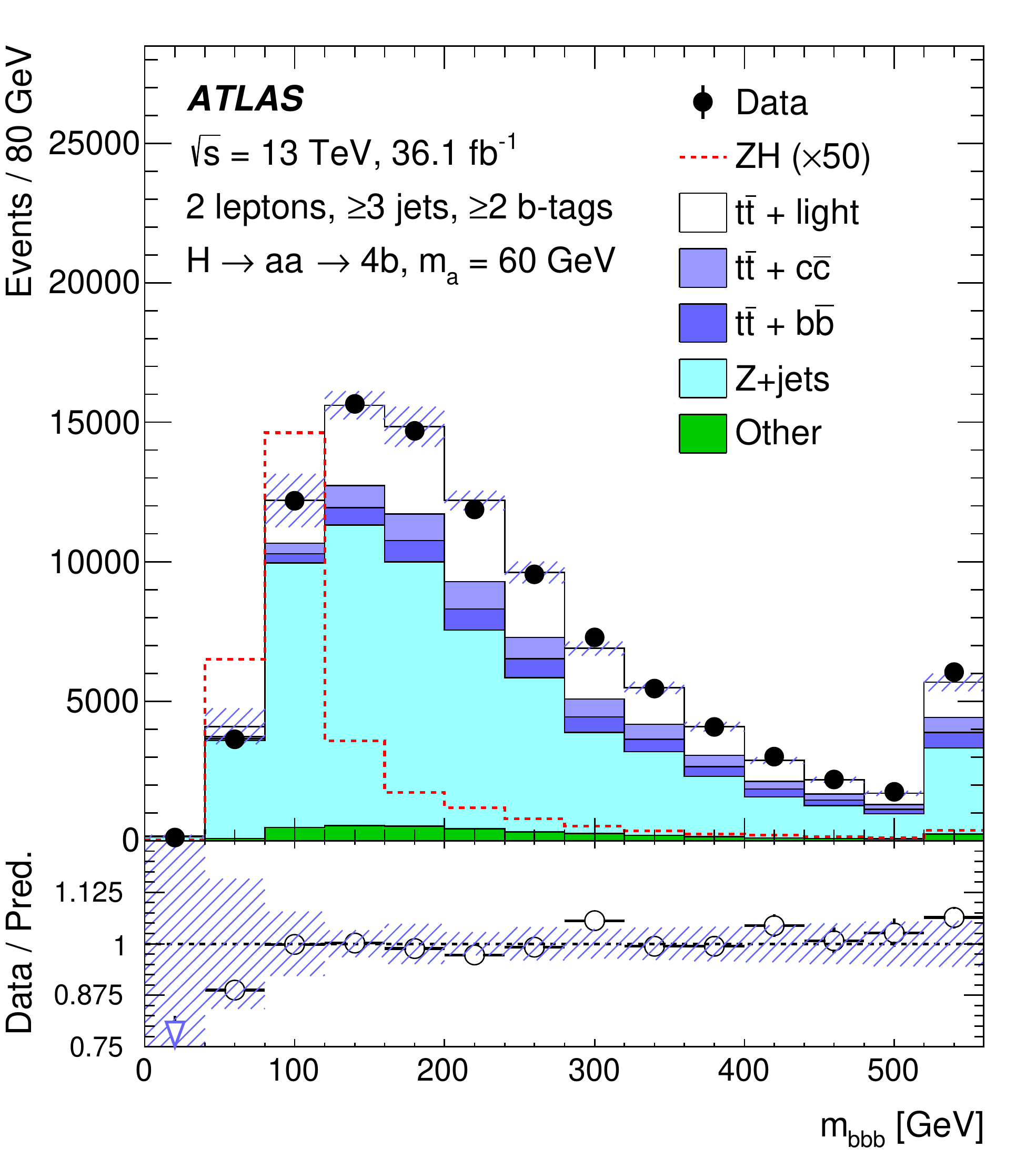}}%
        \subfloat[]{\includegraphics[width=0.33\textwidth]{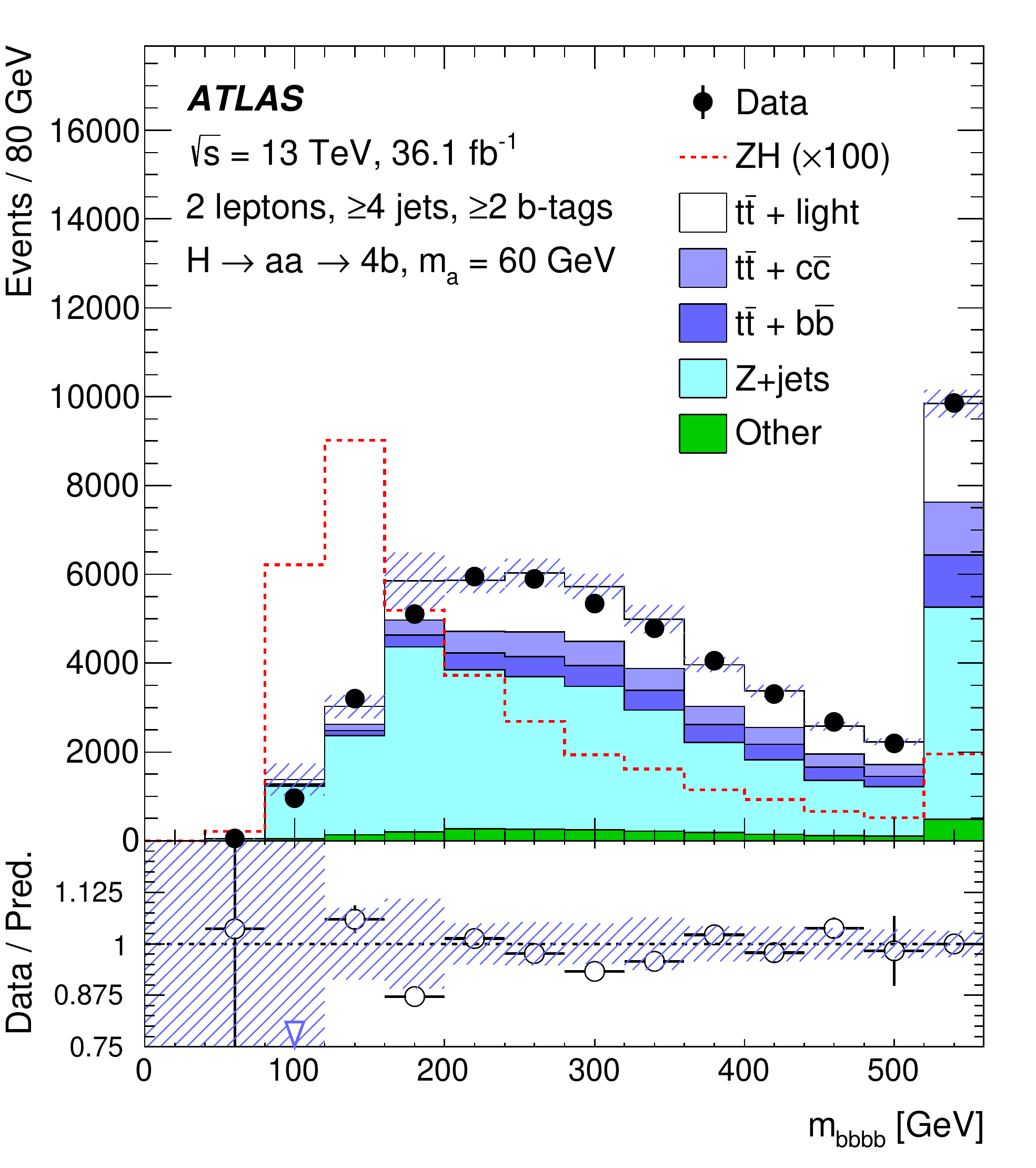}}\\
        \subfloat[]{\includegraphics[width=0.33\textwidth]{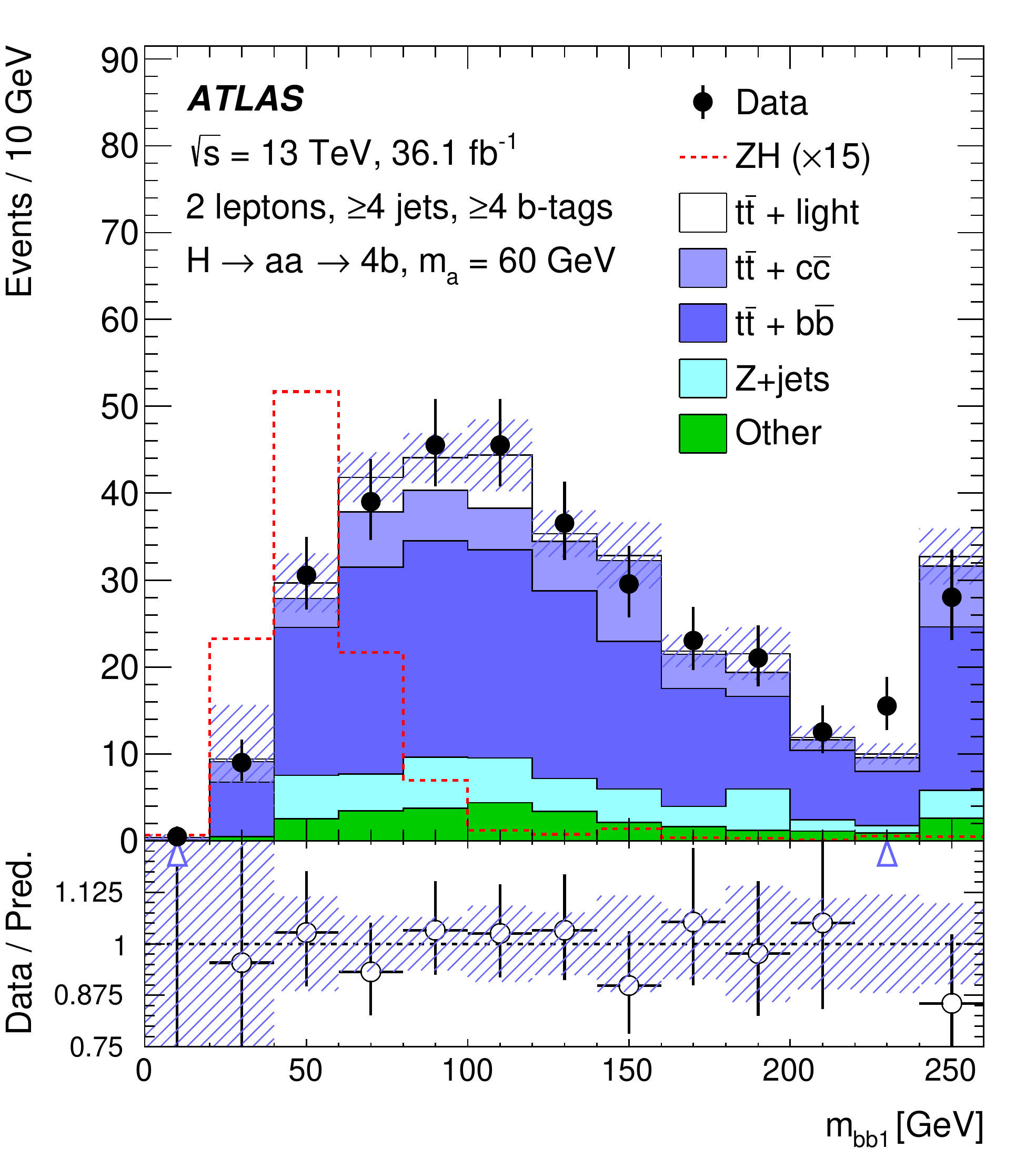}}%
        \subfloat[]{\includegraphics[width=0.33\textwidth]{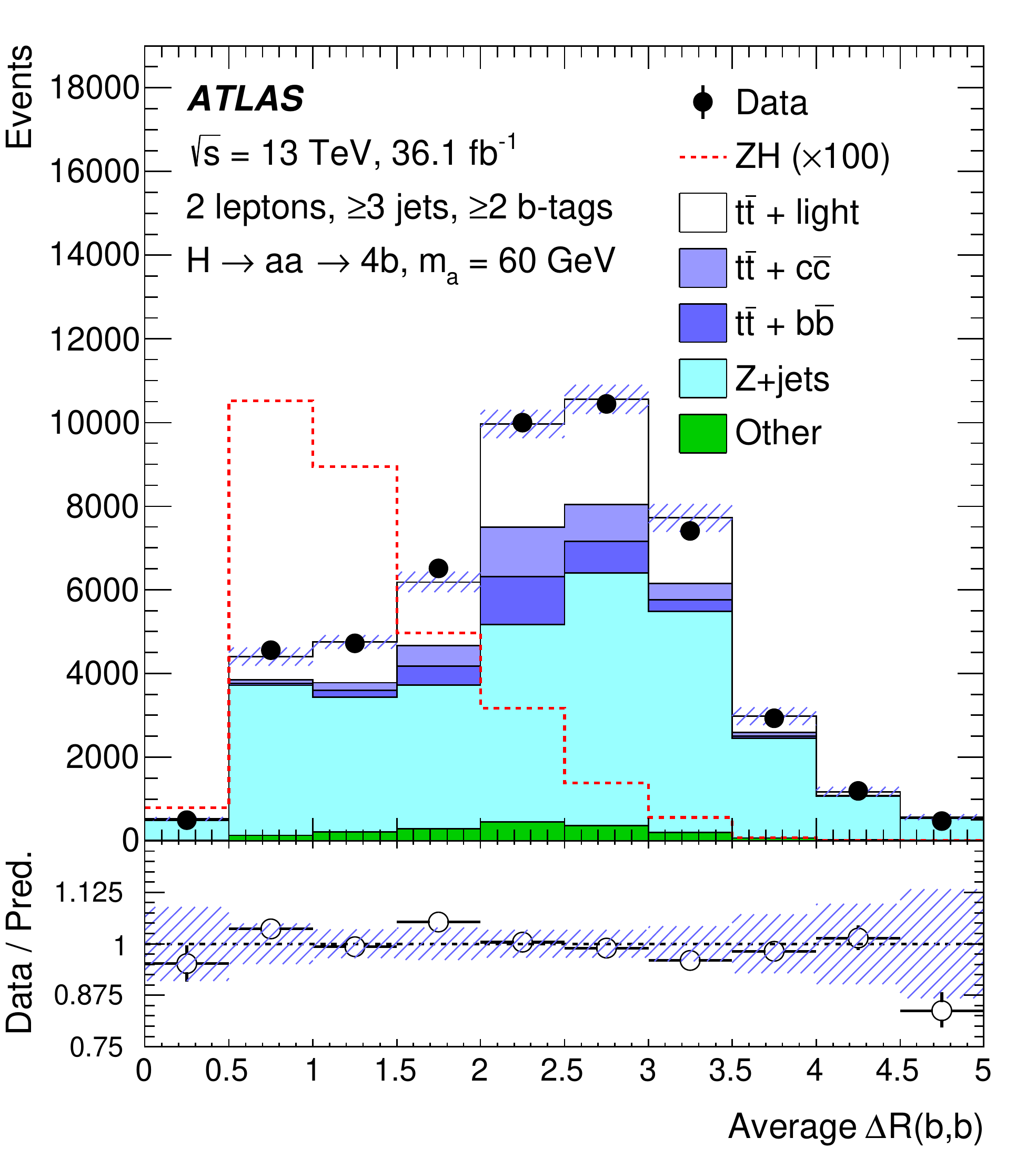}}\\
        \subfloat[]{\includegraphics[width=0.33\textwidth]{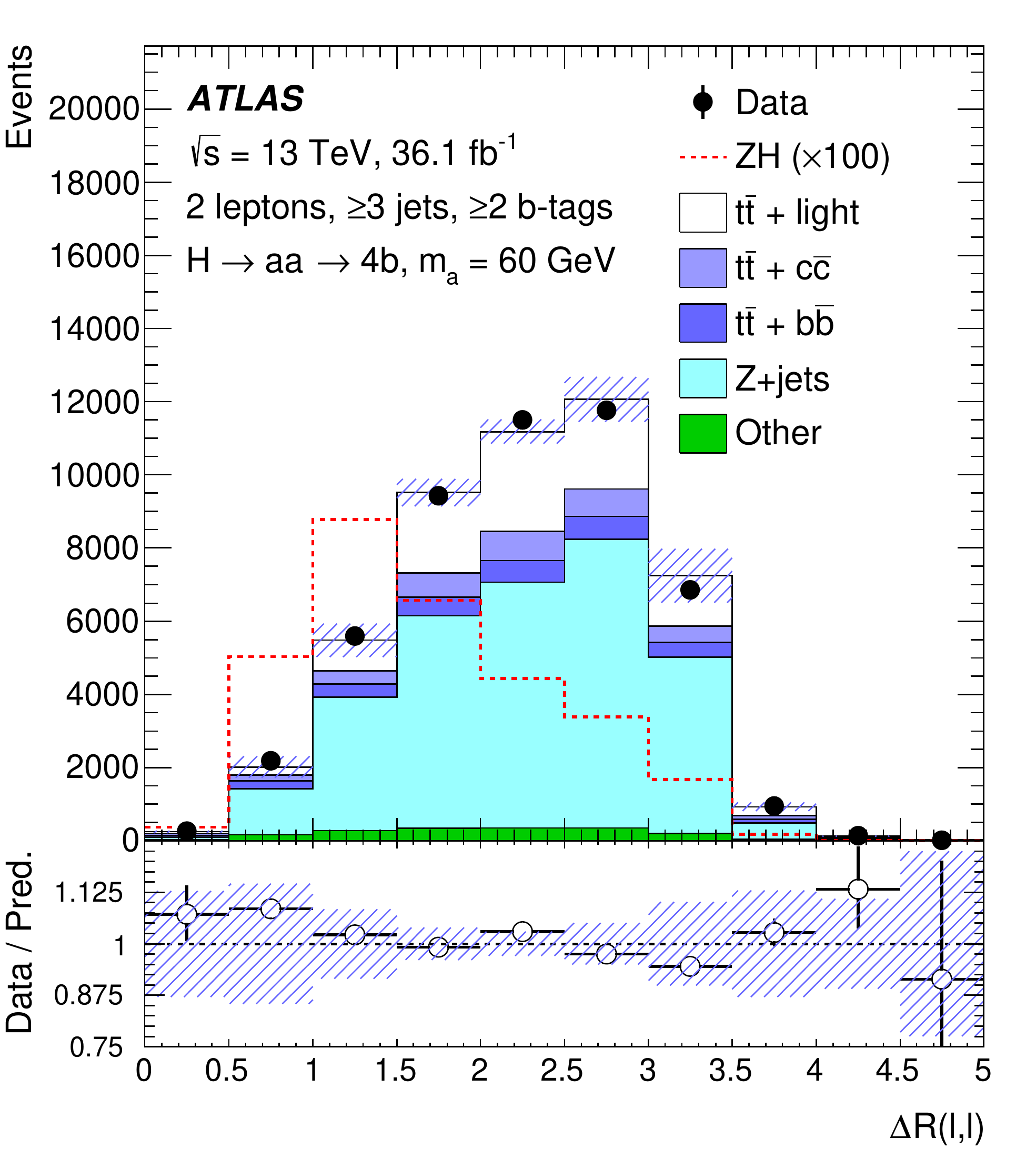}}%
        \subfloat[]{\includegraphics[width=0.33\textwidth]{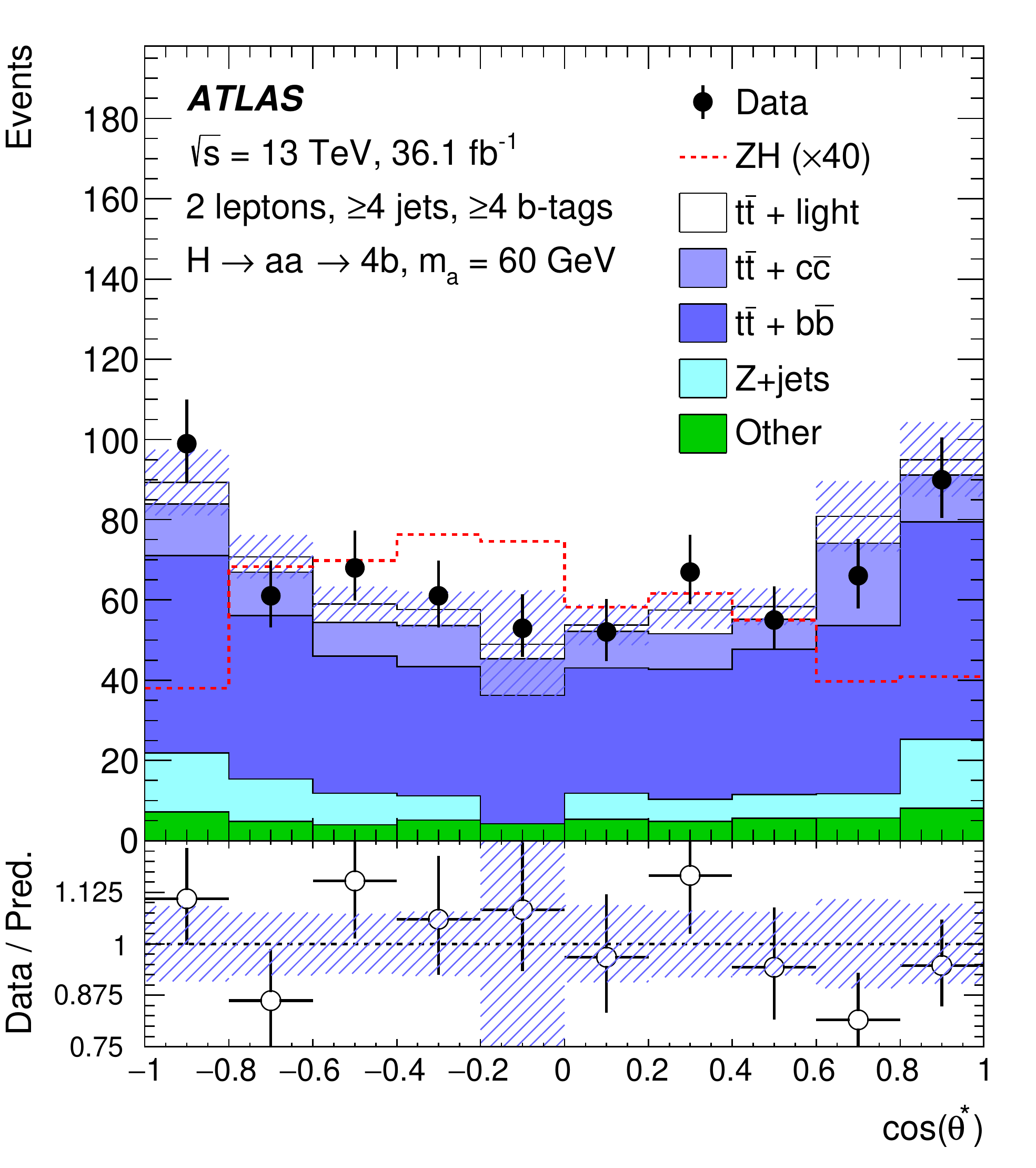}}\\
        \caption{Comparison of data with the  post-fit background
          estimates for (a) $m_{bbb}$, (b)
          $m_{bbbb}$, (c) $m_{bb1}$, (d)
          average $\Delta R$($b$,$b$), (e) $\Delta
          R$($\ell$,$\ell$) and (f) $\cos{\theta^*}$
          in the dilepton sample inclusive in
          number of jets and $b$-tagged jets. Comparisons use events with 
          $\ge3$ jets, except when $\ge4$ jets are necessary to define
          the variable, in which case events with $\ge4$ jets are
          used. Distributions for the signal model
          ($ZH$, $H\rightarrow aa \rightarrow 4b$), with $m_a=$ 60
          GeV, normalized to the SM $pp \rightarrow ZH$ cross-section, assuming $\BR(H\rightarrow aa
          \rightarrow 4b) = 1$ and scaled by a factor as indicated in
          the figure, are overlaid. The hashed area represents the total uncertainty in the
          background. The last bin contains the overflow.}
    \label{fig:kinvar_validation_2}
\end{center}
\end{figure*}

Table~\ref{tab:BDTVars} summarizes the variables used to train
each of the BDT discriminants for the six signal regions. 
Figures~\ref{fig:kinvar_validation_1}~and~\ref{fig:kinvar_validation_2} 
show the expected distributions of the kinematical variables
inclusively in number of jets and $b$-tagged jets. The jets with the
largest values of the $b$-tagging discriminant are used to define the
variables shown. The distributions are obtained ``post-fit'', after
accounting for the systematic uncertainties and applying the
statistical procedure described in sections~\ref{sec:systematics}
and~\ref{sec:results}, respectively.

\FloatBarrier

\section{Systematic uncertainties}
\label{sec:systematics}

Many sources of systematic uncertainties affect this search, including
those related to the integrated luminosity, 
to the reconstruction and identification of leptons and jets, 
and to the modelling of signal and background processes. 
Some uncertainties affect only the overall normalization of the samples, while others also impact
the shapes of the distributions used to categorize events and build the final discriminants.

A single nuisance parameter is assigned to each source of
systematic uncertainty, as described in section~\ref{sec:results}. 
Some of the systematic uncertainties, in particular most of the experimental uncertainties, 
are decomposed into several independent sources, as specified in the following. 
Each individual source then has a correlated effect across all 
channels, analysis categories, signal and background samples.
For modelling uncertainties, especially the \ttbar{} and \Zjets{}
modelling, additional nuisance parameters are included to split some
uncertainties into several sources affecting different
subcomponents of a particular process independently.

The uncertainty of the combined integrated luminosity for 2015 and 2016 is
2.1\%. It is determined using a methodology similar to that detailed
in ref.~\cite{DAPR-2013-01}. Uncertainties in the modelling of pile-up are also
estimated, and cover the differences between the predicted and measured
inelastic cross-sections~\cite{STDM-2015-05}.

Uncertainties associated with leptons arise from the trigger,
reconstruction, identification, and isolation efficiencies,
as well as the momentum scale and resolution. 
These are measured in data using leptons in $Z\to \ell^+\ell^-$,
$J/\psi \to \ell^+\ell^-$ and $W\to e\nu$
events~\cite{ElectronID2016,PERF-2015-10} and have only a small
impact on the result. 

Uncertainties associated with jets arise from their reconstruction and
identification efficiencies. These are due to the uncertainty in the jet
energy scale (JES), resolution and the efficiency of the JVT
requirement that is meant to remove jets from pile-up.
The JES and its uncertainty are derived by combining
information from test-beam data, LHC collision data and
simulation~\cite{PERF-2016-04}. Additional uncertainties are also
considered, associated with the jet flavour and pile-up corrections. The
total per-jet uncertainties are 1--6\%, although the effects are
amplified by the large number of jets in the final state. 

The efficiency to correctly tag $b$-jets is measured in data using
dilepton \ttbar\ events~\cite{Aaboud:2018xwy}. The mistag rate for $c$-jets is 
measured in events with $W$ bosons decays into $q\bar{q}$. 
For light jets, it is measured in multi-jet events using jets
containing secondary vertices and tracks with impact parameters 
consistent with a negative lifetime~\cite{PERF-2012-04}. The uncertainty
associated with the $b$-tagging efficiency ranges between
2\% and 10\% depending on the jet \pt. The size of the uncertainties
associated with the mistag rates is 5--20\% for $c$-jets  depending on
the jet \pt\ and 10--50\% for light jets depending on the jet \pt\ and
$\eta$. 

For the long-lived $a$-boson signals, the secondary vertices of $b$-jets are, on
average, further displaced from the primary vertex than those of
$b$-jets from \ttbar\ events.  An additional ``displaced
$b$-tagging'' systematic uncertainty is applied to long-lived signal
samples to account for a displacement-dependent mismodelling
of the $b$-tagging efficiency. 
The uncertainty is determined using the ``adjusted MC''
method~\cite{ATLAS-CONF-2018-006}, which was originally developed for
the calibration of the mistag rate for light-flavour jets. 
The resulting uncertainty increases approximately linearly with the $a$-boson
proper lifetime, from $\sim\! 2\%$ for prompt $a$-bosons to $\sim\!
10\%$ for proper lifetimes of 10~mm. It is applied in addition to the
standard $b$-tagging uncertainties.

Uncertainties associated with energy scales and resolutions of leptons and jets 
are propagated to $\met$. An uncertainty in the contribution from
charged-particle tracks not associated with reconstructed leptons and jets is also
included in the $\met$ uncertainty~\cite{Aaboud:2018tkc}. 

Several sources of systematic uncertainty affecting the modelling of
the main backgrounds, $\ttbar$ and \Zjets{} are considered. For the
$\ttbar$ background, the procedure closely follows that described in
ref.~\cite{Aaboud:2017rss}.  An uncertainty of 6\% is
assumed for the $\ttbar$ production cross-section~\cite{Czakon:2011xx}, including contributions from
variations of the factorization and renormalization scales, and
uncertainties arising from the PDFs, $\alpha_{\mathrm{S}}$, and the
top-quark mass.

Systematic uncertainties affecting the shape of the $\ttbar$
background account for the choice of generator, the choice of PS and 
hadronization models, and the effects of initial- and
final-state radiation. The uncertainties are derived from comparisons between the 
nominal simulation ({\powheg}+{\pythia}) and alternative samples
produced with \textsc{Sherpa+OpenLoops} (varying the NLO event
generator, PS and hadronization models) or
{\powheg}+\herwigseven~\cite{Bellm:2015jjp} (varying only the PS and 
hadronization models). 

Additional uncertainties are evaluated to account for the use of
{\sherpa}+{\textsc{OpenLoops}} NLO to model the $\ttbb$ and $\ttcc$
backgrounds. Uncertainties are also assessed for the choice of scheme
to treat massive quarks and the choice of PDF sets, as well as the choice
of shower recoil model and scale. Uncertainties are also included to
account for differences in the relative contributions of the
$\ttbb$, $\ttcc$ and $\ttlight$
processes. All uncertainties are treated as uncorrelated across
the $\ttbar$ flavour components. The normalization of the $\ttbb$
process is included as an independent free-floating factor, while the
$\ttcc$ component is assigned a 50\% normalization uncertainty,
derived from studies of alternative background samples~\cite{Aad:2015gra}.

In the case of the \Wjets{} and \Zjets{} backgrounds, all normalizations 
are included as independent free-floating factors. In
the case of the \Zjets{} background in the dilepton channel, a
separate normalization factor is considered for each jet and
$b$-tagged jet multiplicity bin: (2$\ell$, $\geq$3j, 2b), (2$\ell$, 3j,
3b), (2$\ell$, $\geq$4j, 3b) and (2$\ell$, $\geq$4j, $\geq$4b). 
Additional assigned uncertainties are based on variations of the
factorization and renormalization scales and of the matching
parameters in the \textsc{Sherpa} simulation.

A cross-section uncertainty of $^{+5\%}_{-4\%}$ is assigned to the
three single-top-quark production 
modes~\cite{Kidonakis:2011wy,Kidonakis:2010ux,Kidonakis:2010tc}. 
For the $Wt$ and $t$-channel production modes, 
uncertainties associated with the choice of PS and
hadronization model and with initial- and final-state radiation
are evaluated by using a set of alternative samples. 
The uncertainty in modelling of the interference between $Wt$ and \ttbar\ 
production at NLO is assessed by comparing the default simulation to
an alternative one that resolves the interference at the cross-section
level (``diagram subtraction'' scheme) instead of the amplitude level 
(``diagram removal'' scheme)~\cite{Frixione:2008yi}.

A 50\% normalization uncertainty in the diboson background is assumed,
which includes uncertainties in the inclusive cross-section and the
production of additional jets~\cite{multibosonMC}. The
uncertainties in the $\ttbar W$ and $\ttbar Z$ NLO cross-section
predictions are 13\% and 12\%,
respectively~\cite{deFlorian:2016spz,Campbell:2012dh}, due to PDF and scale
uncertainties, and are treated as uncorrelated
between the two processes. An additional modelling uncertainty for
$\ttbar W$ and $\ttbar Z$, related to the choice of event generator,
PS and hadronization models, is derived from comparisons of the
nominal samples with alternative ones generated with \textsc{Sherpa}.

In the single-lepton channel, uncertainties in the estimation of the
background with fake or non-prompt leptons come from the limited number of events in the
data sample without the lepton isolation requirement and from uncertainties
in the measured non-prompt and prompt lepton efficiencies. The
normalization uncertainty assigned to this background is 50\%, as
derived by comparing the background prediction with data in
control regions obtained by inverting the requirements on $\met$
and on $\mtw$. An uncertainty in the shape of the predicted background
distribution covers the difference between the prediction obtained
using an inclusive sample before $b$-tagging is applied and the 
prediction after $b$-tagging. In the dilepton channel, the simulated
non-prompt lepton background is assigned a 25\% uncertainty, 
correlated across lepton flavours and all analysis categories.

Several sources of systematic uncertainty affect the theoretical
modelling of the signal. Uncertainties originate from the
choice of PDFs, the factorization and renormalization scales, and the
PS, hadronization and UE models. 

\section{Results}
\label{sec:results}

The distributions of the discriminant for each of the analysis categories 
are combined in a profile likelihood fit to test for the presence of signal,
while simultaneously determining the normalizations and constraining
the differential distributions of the most important background components.
As described in section~\ref{sec:analysisStrategy}, 
in the signal regions, the output of the BDT classifier is used as the
discriminant, while \HT\ or the invariant
mass of the $b$-jets is used in the control regions.

The likelihood function, ${\cal L}(\mu,\theta)$, is constructed as a
product of Poisson probability terms over all bins in each distribution.
The Poisson probability depends on the predicted number of events in
each bin, which in turn is a function of the signal-strength parameter 
$\mu = \sigma \times \BR(H\rightarrow aa \rightarrow 4b)$, where $\sigma$ are
the $pp \rightarrow WH$ and $ZH$ cross-sections. 
The nuisance parameters, $\theta$, encode the effects of systematic
uncertainties. The nuisance parameters are implemented in the
likelihood function as Gaussian, log-normal or Poisson priors. The
statistical uncertainty of the prediction, which incorporates the
statistical uncertainty of the simulated events and of the data-driven
fake and non-prompt lepton background estimate, is included in the
likelihood as a nuisance parameter for each bin. 

The likelihood function depends on six free-floating normalization
factors for $\ttbb$, $Z+\mathrm{jets}$ for the four jet
and $b$-tagged jet multiplicities and $W+\mathrm{jets}$. No prior
knowledge from theory or subsidiary measurements is assumed for the
normalization factors, hence they are only constrained by the profile
likelihood fit to the data. As shown in
table~\ref{tab:fitresults_normfactors}, the normalization factors are
compatible with SM expectations within the uncertainties. 
The other main background components, particularly $\ttlight$ and 
$\ttcc$, are also compatible with SM expectations within the 
uncertainties described in section~\ref{sec:systematics}. In the
combination of the single-lepton and dilepton channels, the ratio of
$WH$ to $ZH$ cross-sections is assumed to follow the SM prediction. 

\begin{table}[ht!]
\centering
\caption{\label{tab:fitresults_normfactors} Normalization factors
  included as independent free-floating factors in the likelihood
  fit. The uncertainties include statistical and systematic components.}
\begin{tabular}{l|ccc}
\toprule
    & \multicolumn{3}{c}{Normalization factor} \\
    \cline{2-4}
Sample & Single-lepton & Dilepton & Combination \\
\midrule
$\ttbb$ & $1.5 \pm 0.5$ & $0.9 \pm 0.3$  & $1.1 \pm 0.3$ \\
\Wjets{} & $0.7 \pm 0.3$ & -- & $0.7 \pm 0.3$ \\
\Zjets{} (2$\ell$, $\geq$3j, 2b) & -- & $1.0 \pm 0.1$ & $1.1 \pm 0.1$ \\
\Zjets{} (2$\ell$, 3j, 3b) & -- & $1.2 \pm 0.2$ & $1.1 \pm 0.2$ \\
\Zjets{} (2$\ell$, $\geq$4j, 3b) & -- & $1.1 \pm 0.2$ & $1.1 \pm 0.1$ \\
\Zjets{} (2$\ell$, $\geq$4j, $\geq$4b) & -- & $1.4 \pm 0.5$ & $1.4 \pm 0.5$ \\
\bottomrule
\end{tabular}
\end{table}

The test statistic $t_\mu$ is defined as the profile likelihood ratio: 
$t_\mu = -2\ln({\cal L}(\mu,\hat{\hat{\theta}}_\mu)/{\cal L}(\hat{\mu},\hat{\theta}))$,
where $\hat{\mu}$ and $\hat{\theta}$ are the values of the parameters
which maximize the likelihood function, and $\hat{\hat{\theta}}_\mu$
are the values of the nuisance parameters which maximize the
likelihood function for a given value of $\mu$. 
This test statistic is used to measure the probability that the
observed data is compatible with the signal+background hypothesis, and
to perform statistical inferences about $\mu$, 
such as upper limits using the CL$_{\textrm{s}}$ method~\cite{Cowan:2010js,Read:2002hq,Junk:1999kv}. 
The uncertainty of the best-fit value of the signal strength,
$\hat{\mu}$, is obtained when varying $t_\mu$ by one unit. 

After performing the fit, the leading sources of systematic
uncertainty are the modelling of the $\ttbar$ and \Zjets{} backgrounds and the
$b$-, $c$- and light-jet-tagging efficiencies. Table~\ref{tab:SystSummary_H4b}
summarizes the main systematic uncertainties by indicating their impact on 
the normalization of the main backgrounds and the signal ($\hat{\mu}$)
with $m_a$ = 60~\GeV{} in the most sensitive signal regions (1$\ell$,
4j, 4b) and (2$\ell$, $\geq$4j, $\geq$4b). The uncertainties in the
normalization are obtained by varying the parameter corresponding to
the source of uncertainty under consideration up and down by one standard
deviation, while keeping the other nuisance parameters fixed at their central value. 

\begin{table}[htbp]
\begin{center}
  \caption{Summary of the impact of the considered systematic
  uncertainties (in \%) on the yields for the main backgrounds and the
  signal ($m_a$ = 60~\GeV{}) for the single-lepton and dilepton
  regions (1$\ell$, 4j, 4b) and (2$\ell$, $\geq$4j, $\geq$4b) after the fit.
  The total uncertainty can differ from the sum in quadrature of
  individual sources due to correlations.
  }
  \label{tab:SystSummary_H4b}
\begin{small}
  \begin{tabular*}{\textwidth}{@{\extracolsep{\fill}}l|cccc|ccc}
\toprule
    & \multicolumn{7}{c}{Impact on yield [\%]} \\
    \cline{2-8}
    & \multicolumn{4}{c|}{Single lepton} & \multicolumn{3}{c}{Dilepton}\\
    Systematic uncertainty & $WH$ signal & $t\bar{t}$ + light &
                                         $t\bar{t} + c\bar{c}$ & $t\bar{t} + b\bar{b}$ & $ZH$ signal & $t\bar{t} + b\bar{b}$ & \Zjets \\ 
\midrule
  Luminosity                   & 2                  & 2                  & 2                  & 2                  & 2                  & 2                  & 2                 \\
  Lepton efficiencies          & 1                  & 1                  & 1                  & 1                  & 1                  & 1                  & 1                  \\
  Jet efficiencies             & 1                  & 1                  & 1                  & 1                  & 1                  & 1                  & 2                  \\
  Jet energy resolution        & 5                  & 4                  & 4                  & 1                  & 7                  & 5                  & 6                  \\
  Jet energy scale             & 4                  & 2                  & 3                  & 2                  & 4                  & 4                  & 7                  \\
  $b$-tagging efficiency       & 16                 & 5                  & 4                  & 9                  & 20                 & 14                 & 17                 \\
  $c$-tagging efficiency       & 1                  & 5                  & 9                  & 3                  & 7                  & 1                  & 1                  \\
  Light-jet-tagging efficiency & 2                  & 16                 & 5                  & 2                  & 1                  & 3                  & 1                  \\
  Theoretical cross-sections   & --                 & 5                  & 5                  & 5                  & --                 & 8                  & --                 \\
  $t\bar{t}$: modelling        & --                 & 5                  & 35                 & 45                 & --                 & 19                 & --                 \\ 
  $t\bar{t}$+HF: normalization & --                 & --                 & 31                 & 33                 & --                 & 38                 & --                 \\
  $t\bar{t}$+HF: modelling     & --                 & --                 & 10                 & 5                  & --                 & 7                  & --                 \\
  \Zjets{}: normalization      & --                 & --                 & --                 & --                 & --                 & --                 & 38                 \\
  Signal modelling             & 7                  & --                 & --                 & --                 & 10                 & --                  & --                 \\
  Displaced $b$-tagging  & 5--8                  & --                 & --                 & --                 & 5--8                 & --                  & --                 \\
  \hline
  Total                        & 33--34                 & 32                 & 75                 & 58                 & 30--31                 & 32                 & 36                 \\
\bottomrule
  \end{tabular*}
  \end{small}
  \end{center}
\end{table}

\begin{figure*}[htbp]
\begin{center}
\subfloat[]{\includegraphics[width=0.33\textwidth]{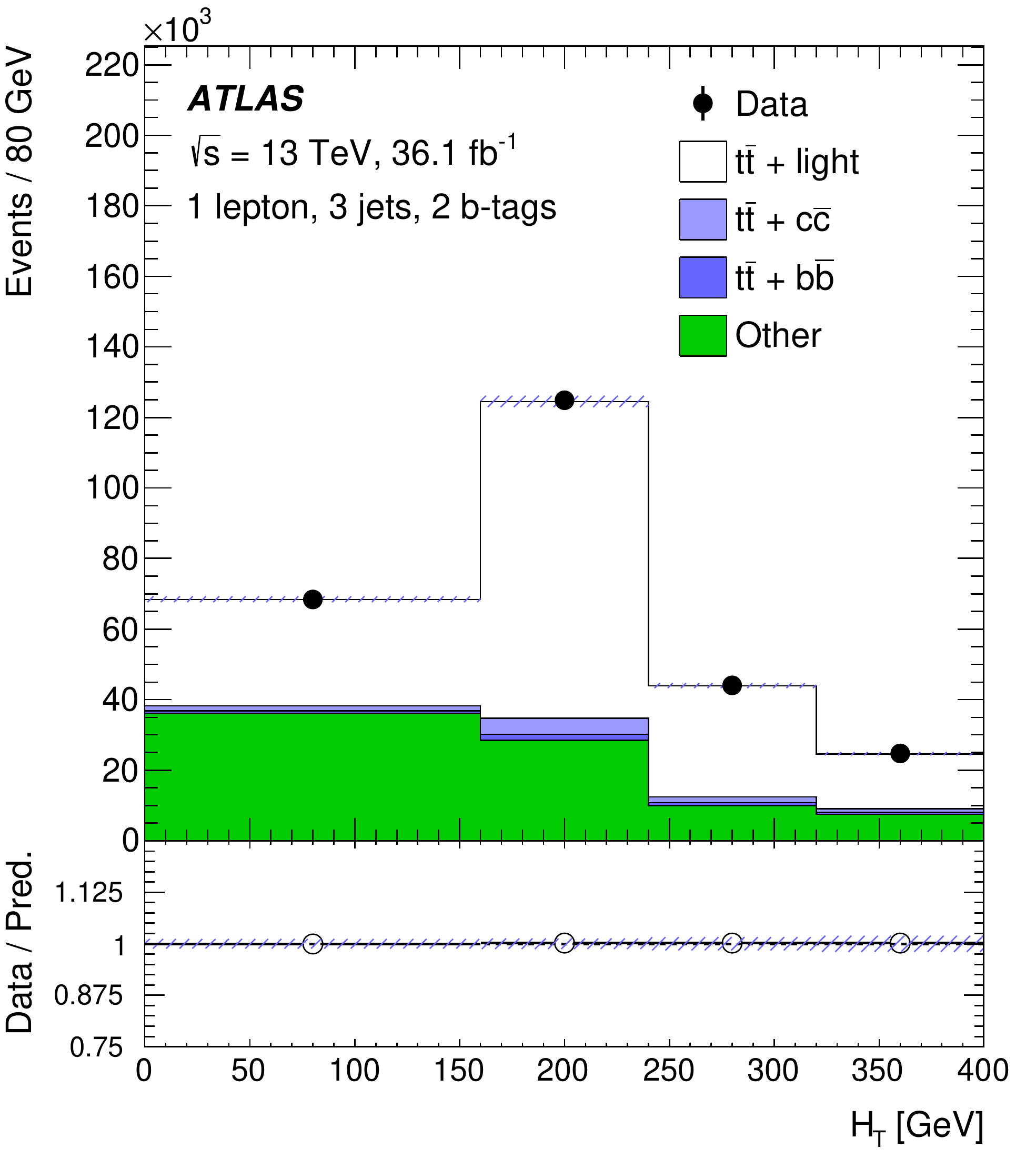}}%
\subfloat[]{\includegraphics[width=0.33\textwidth]{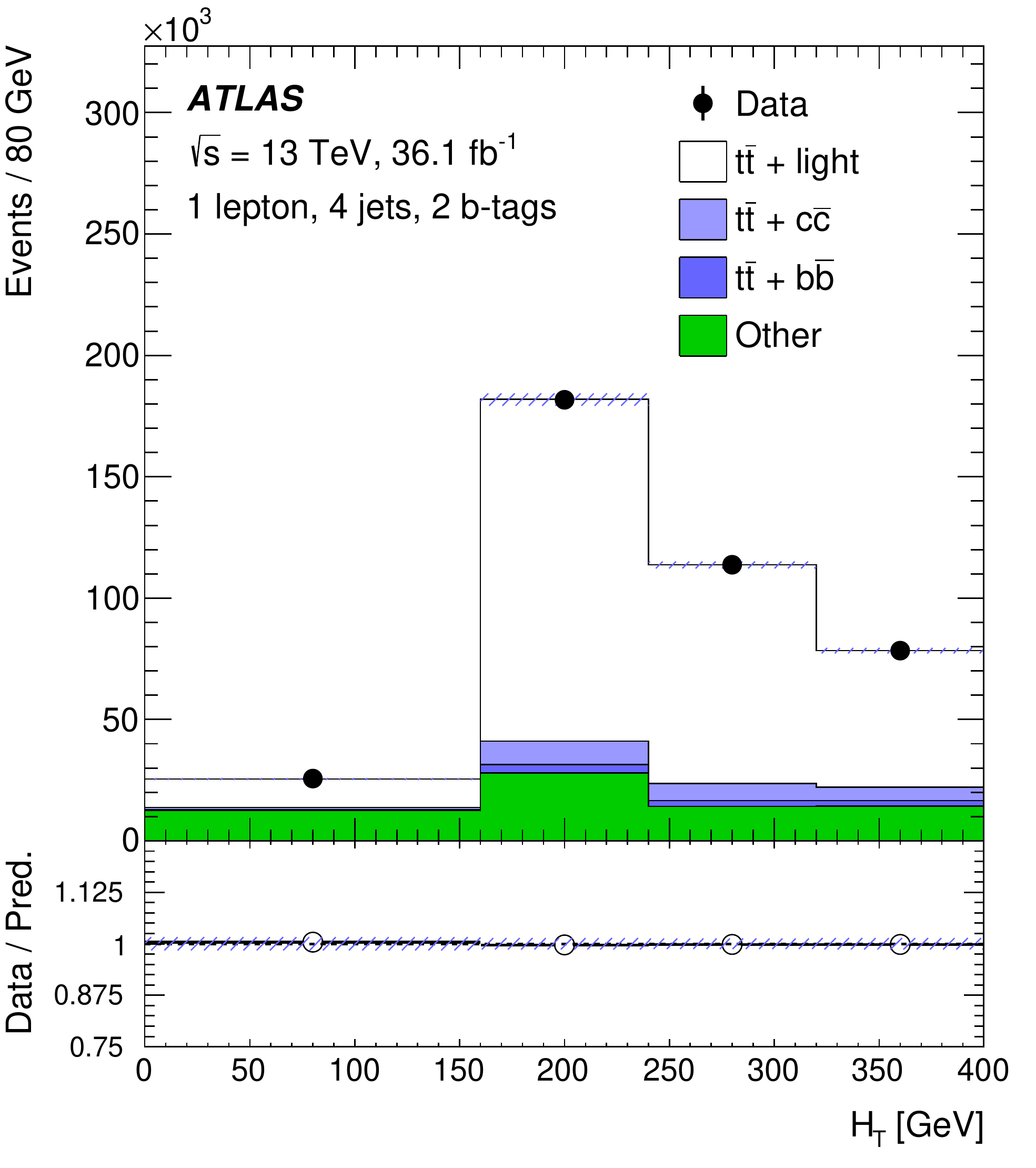}}\\
\subfloat[]{\includegraphics[width=0.33\textwidth]{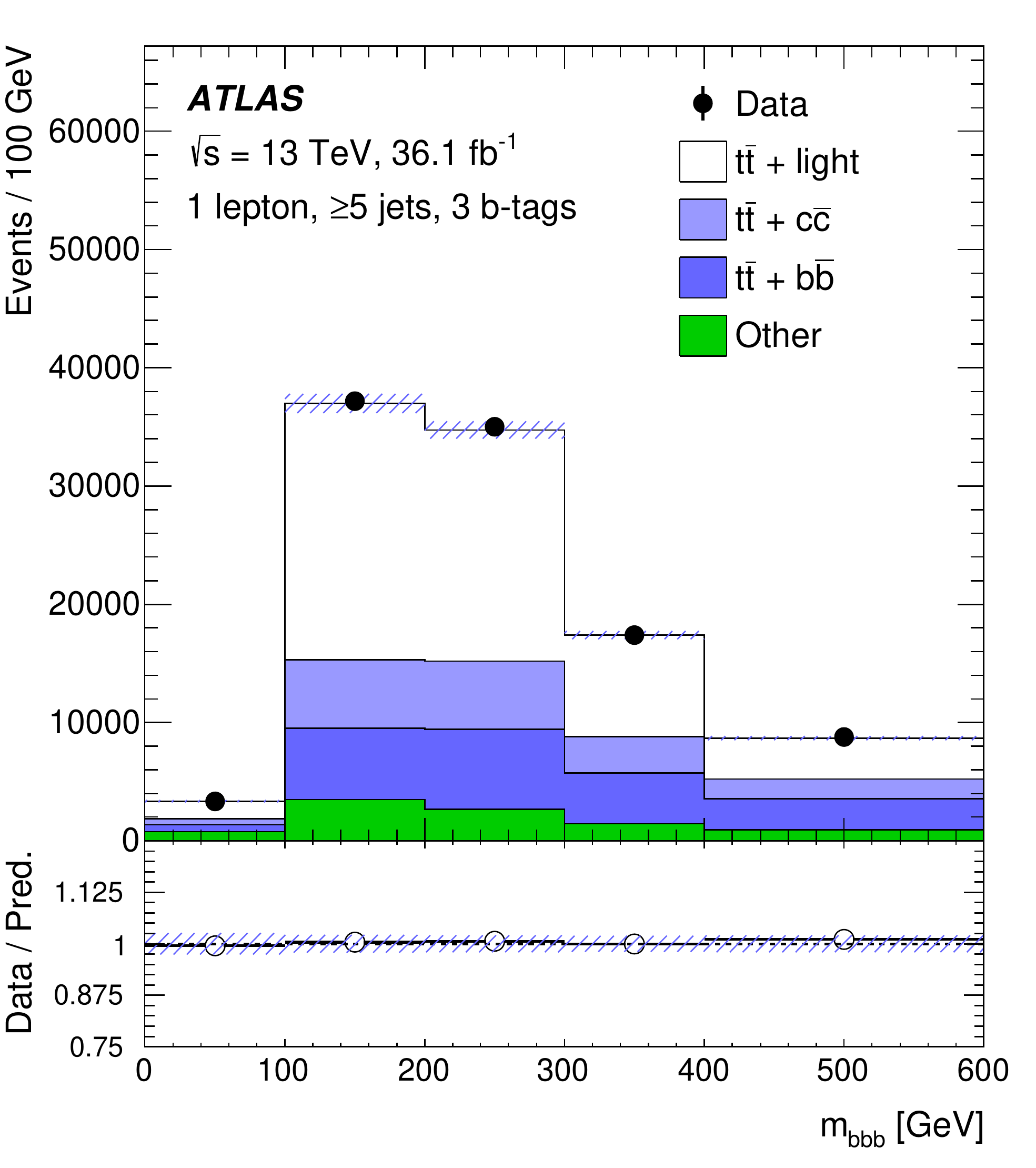}}%
\subfloat[]{\includegraphics[width=0.33\textwidth]{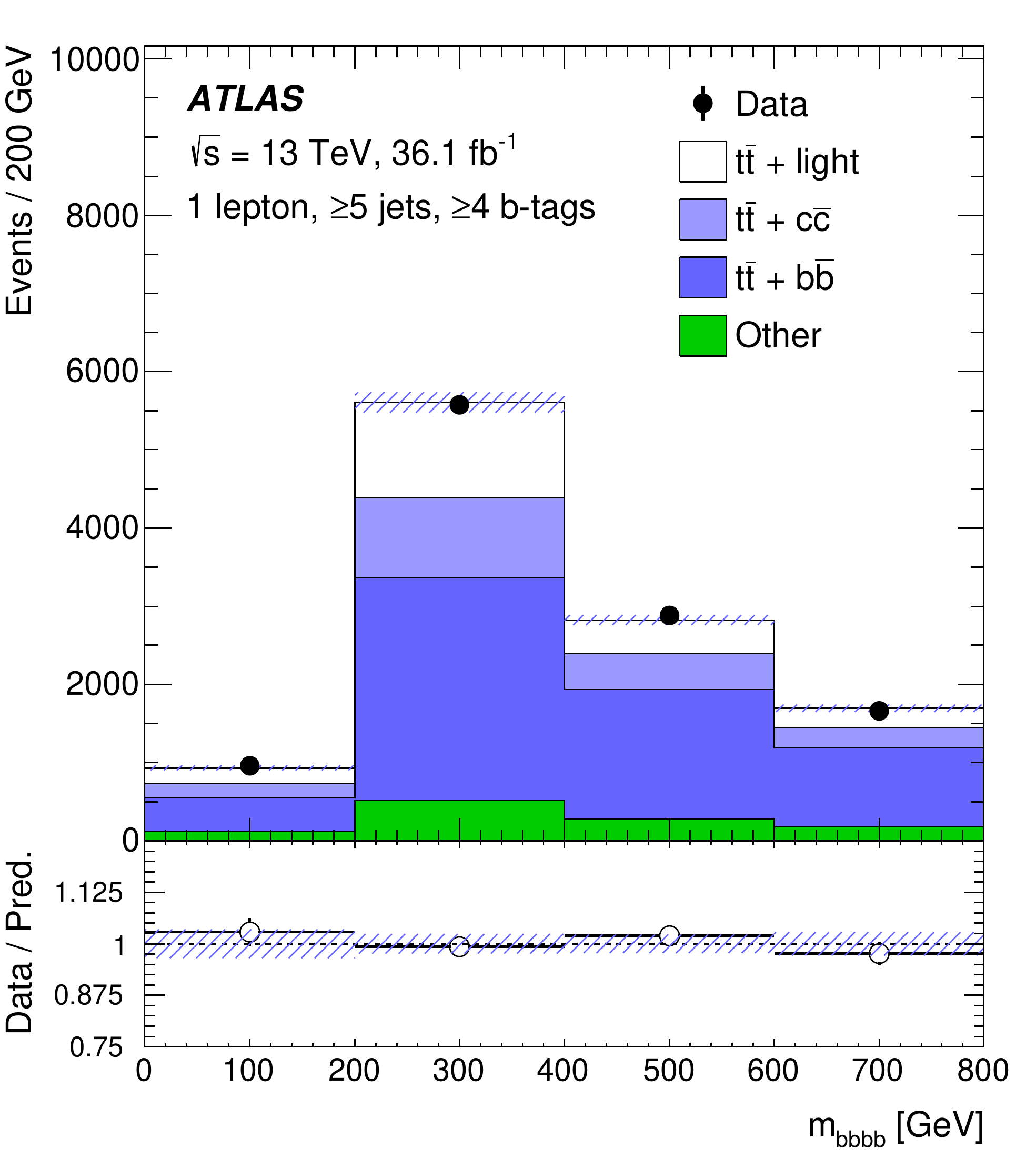}}\\
\caption{Comparison between data and prediction in the single-lepton
  control regions of the $\HT$ variable for (a) (1$\ell$, 3j, 2b) and
  (b) (1$\ell$, 4j, 2b) and the invariant mass of $b$-tagged jets for
  (c) (1$\ell$, $\geq$5j, 3b) and (d) (1$\ell$, $\geq$5j, $\geq$4b),
  after the combined single-lepton and dilepton fit to the data. The
  last bin contains the overflow. 
}
\label{fig:CR_sl}
\end{center}
\end{figure*}

\begin{figure*}[htbp]
\begin{center}
\subfloat[]{\includegraphics[width=0.33\textwidth]{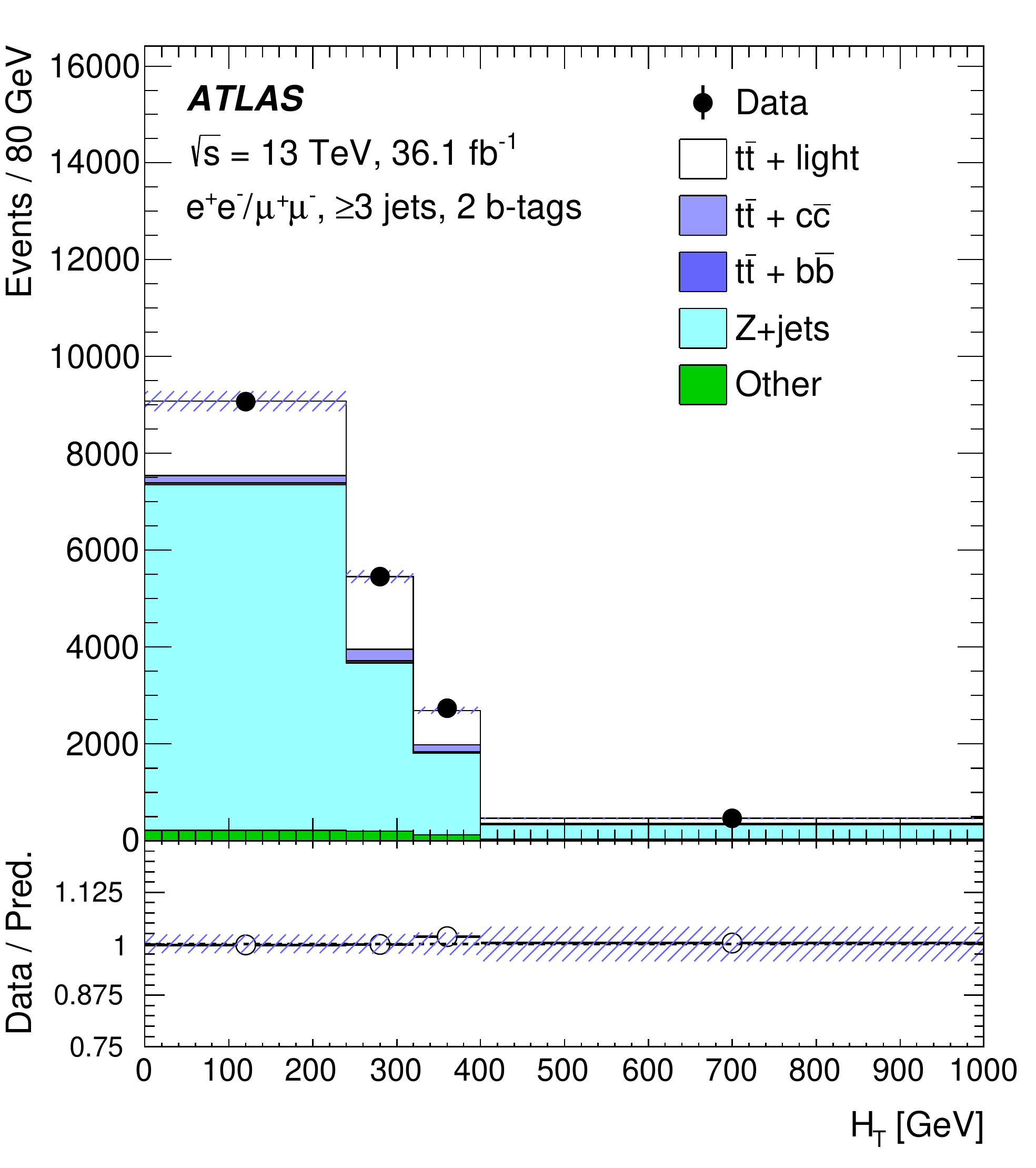}}%
\subfloat[]{\includegraphics[width=0.33\textwidth]{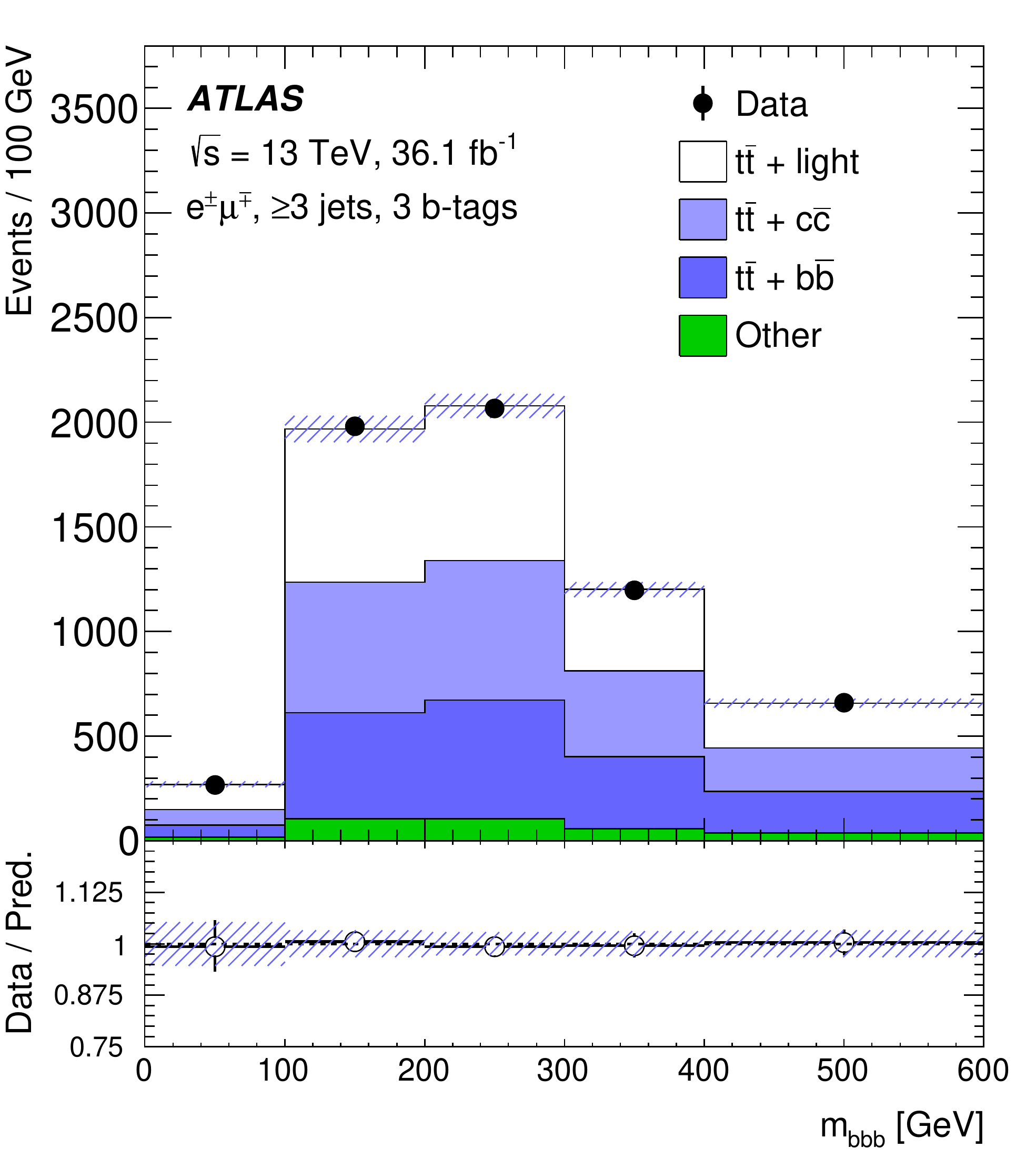}}%
\subfloat[]{\includegraphics[width=0.33\textwidth]{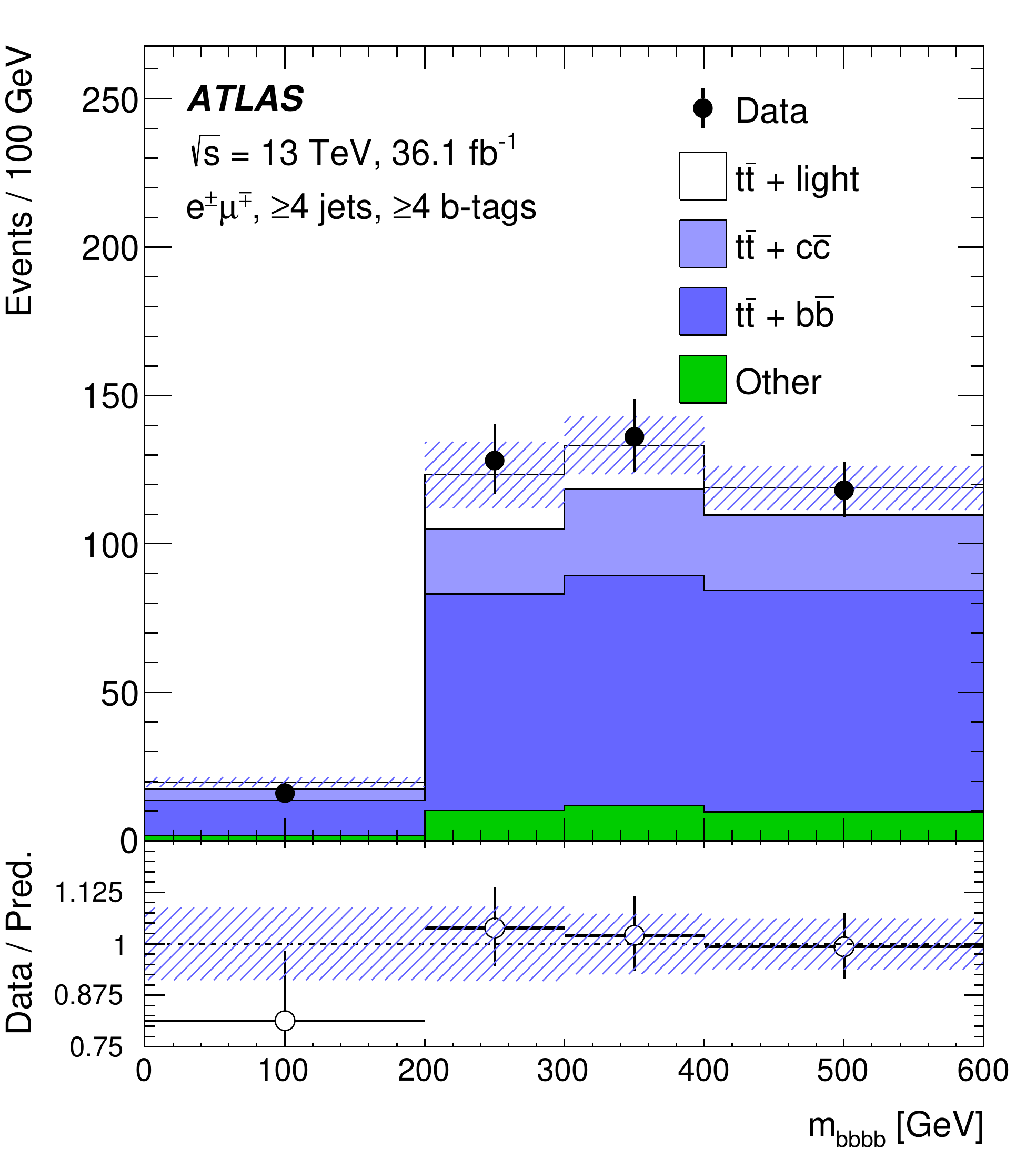}}\\
\caption{Comparison between data and prediction in the dilepton
  control regions of (a) the $\HT$ variable for same-flavour dilepton
  events with (2$\ell$, 3j, 2b) and the invariant mass of $b$-tagged
  jets for different-flavour dilepton events with (b) (2$\ell$,
  $\geq$3j, 3b) and (c) (2$\ell$, $\geq$4j, $\geq$4b), after the
  combined single-lepton and dilepton fit to the data. The last bin
  contains the overflow. 
}
\label{fig:CR_dl}
\end{center}
\end{figure*}

\begin{figure*}[htbp]
\begin{center}
\subfloat[]{\includegraphics[width=0.33\textwidth]{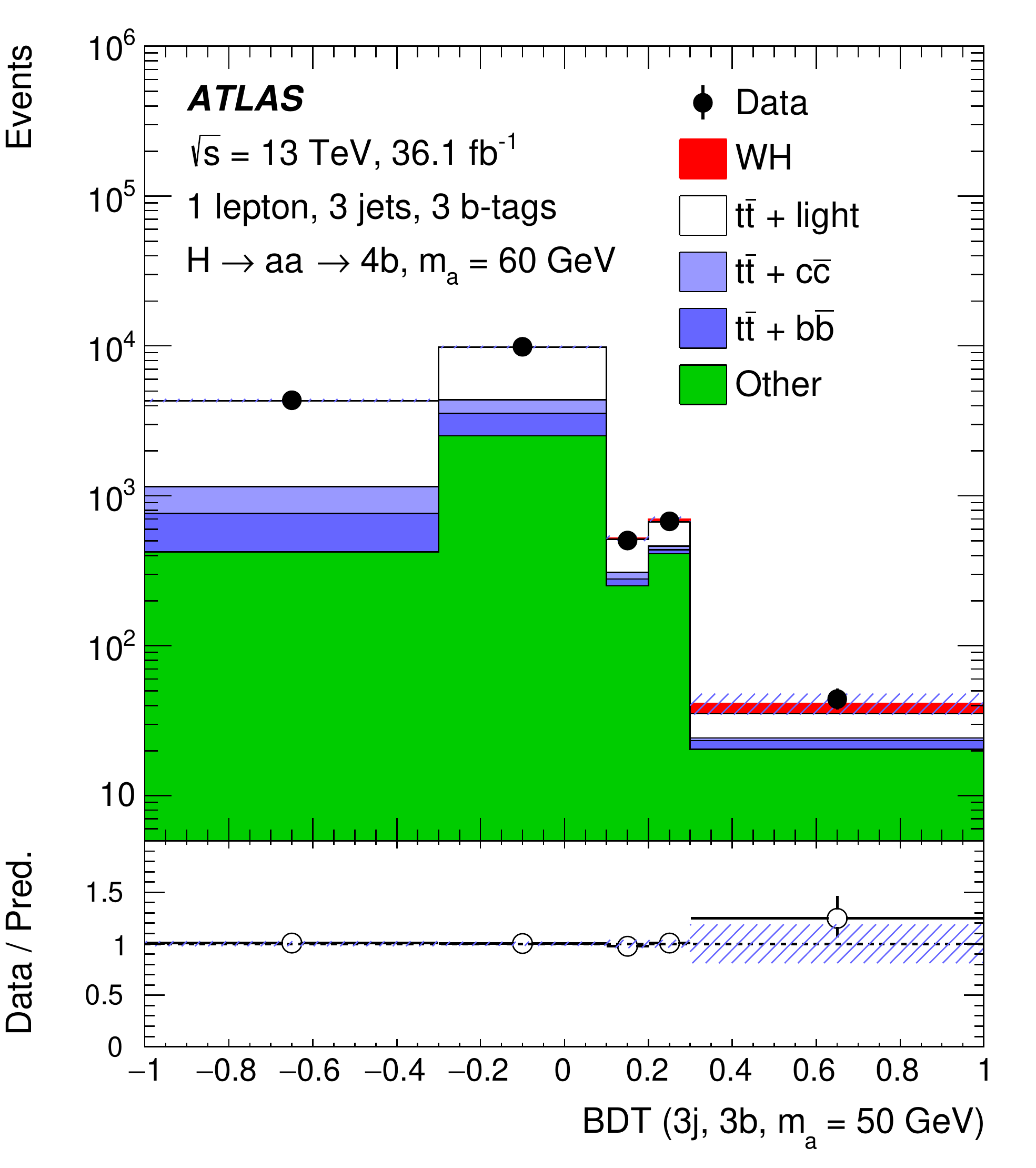}}%
\subfloat[]{\includegraphics[width=0.33\textwidth]{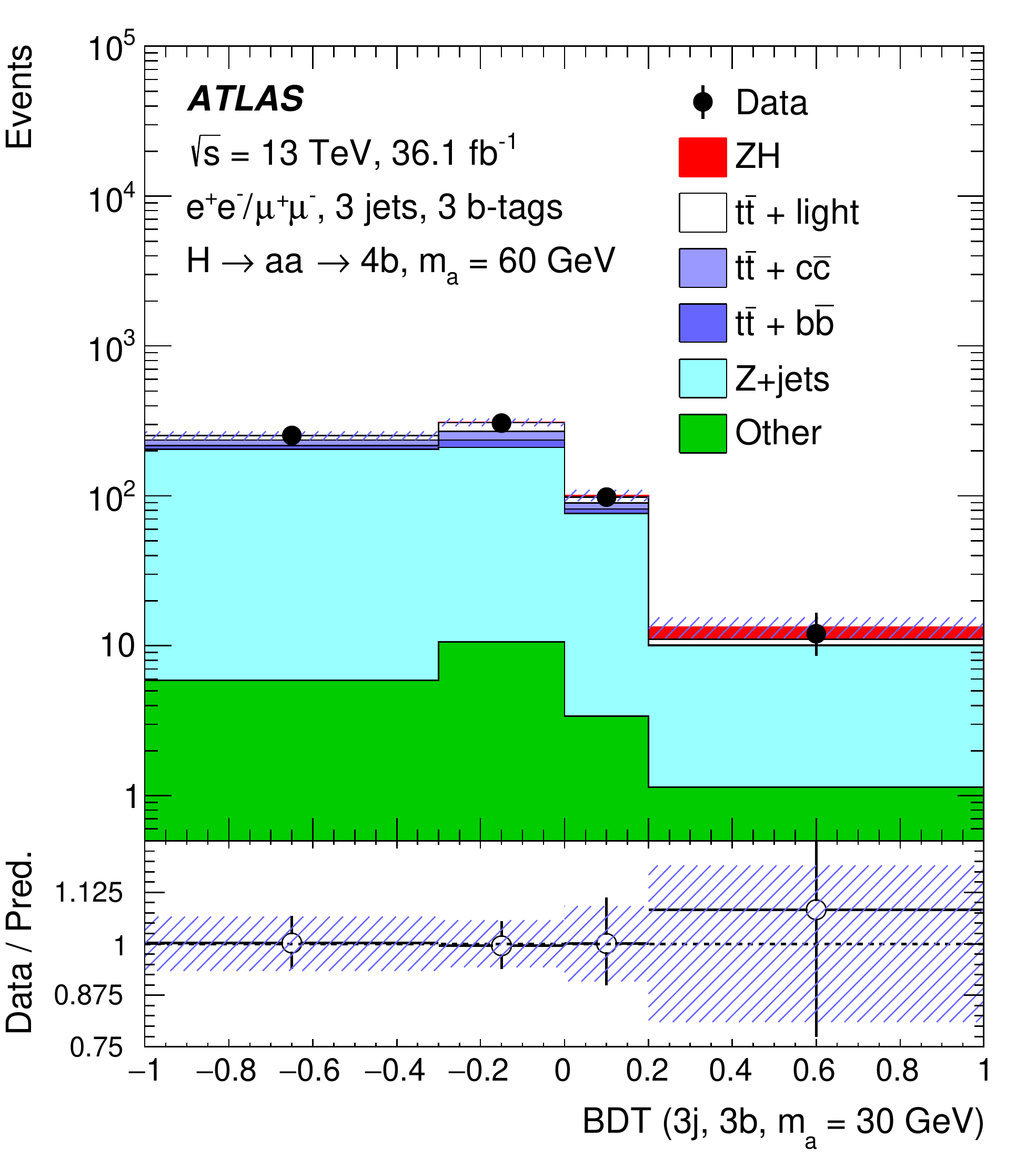}}\\
\subfloat[]{\includegraphics[width=0.33\textwidth]{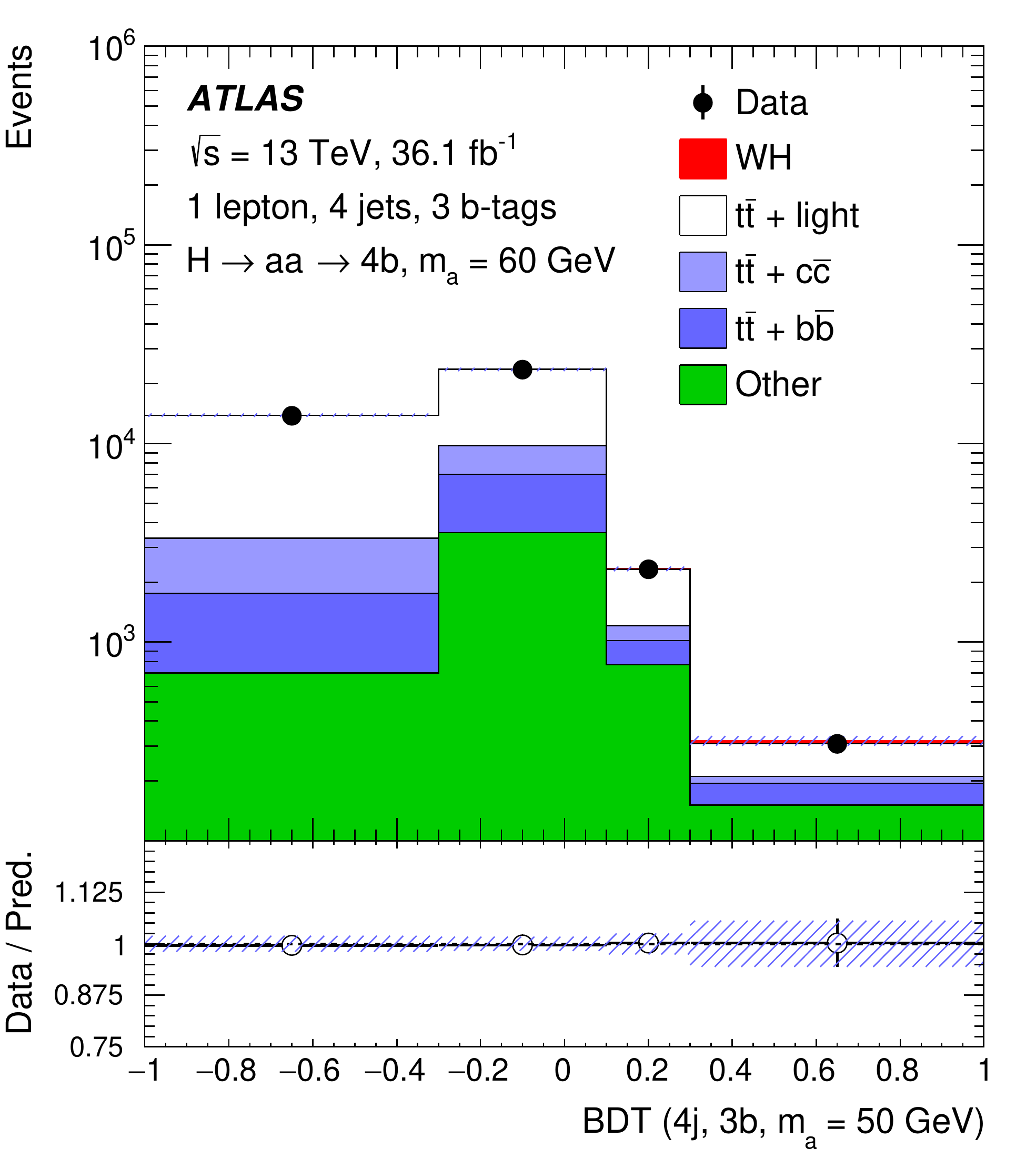}}%
\subfloat[]{\includegraphics[width=0.33\textwidth]{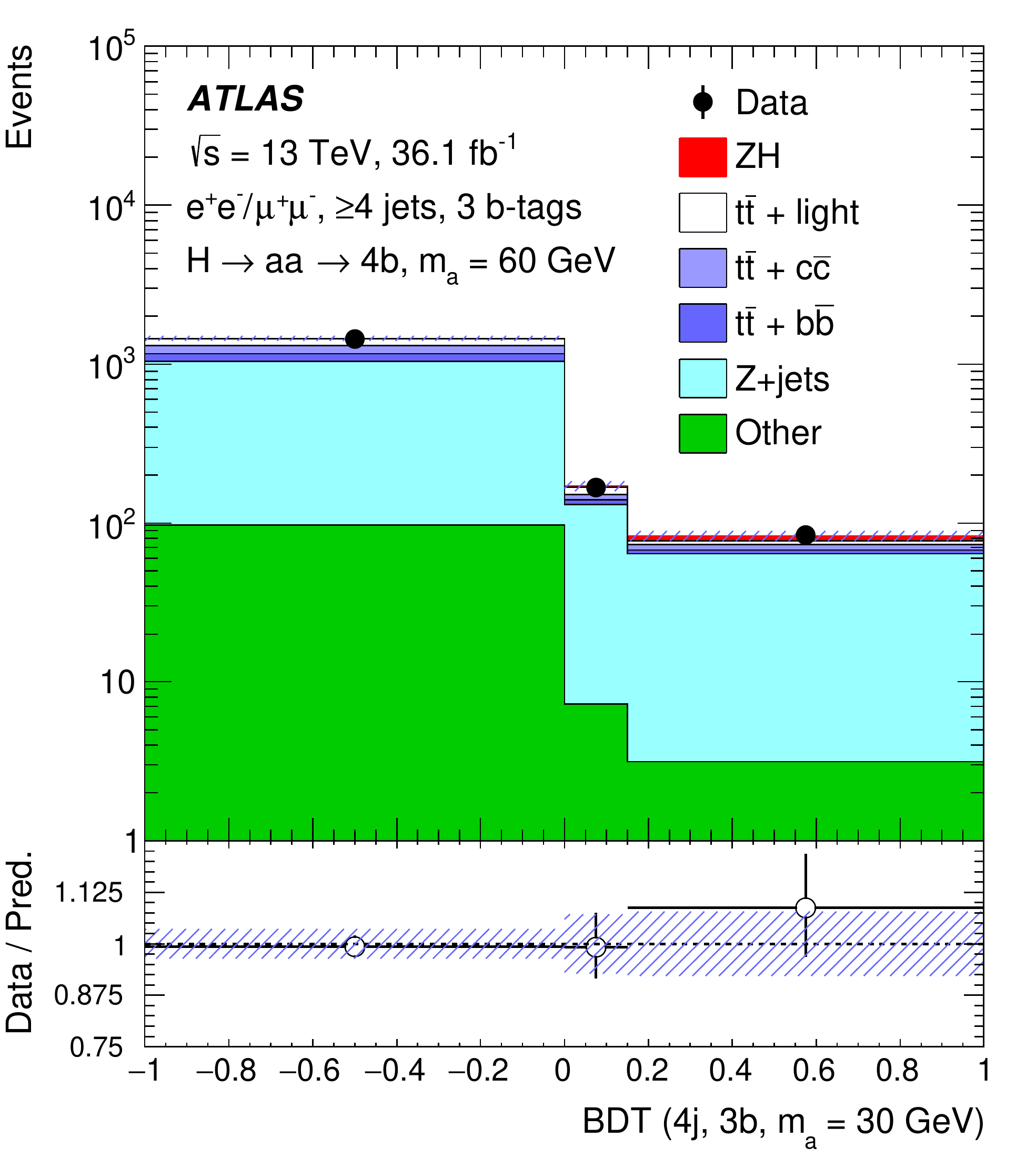}}\\
\subfloat[]{\includegraphics[width=0.33\textwidth]{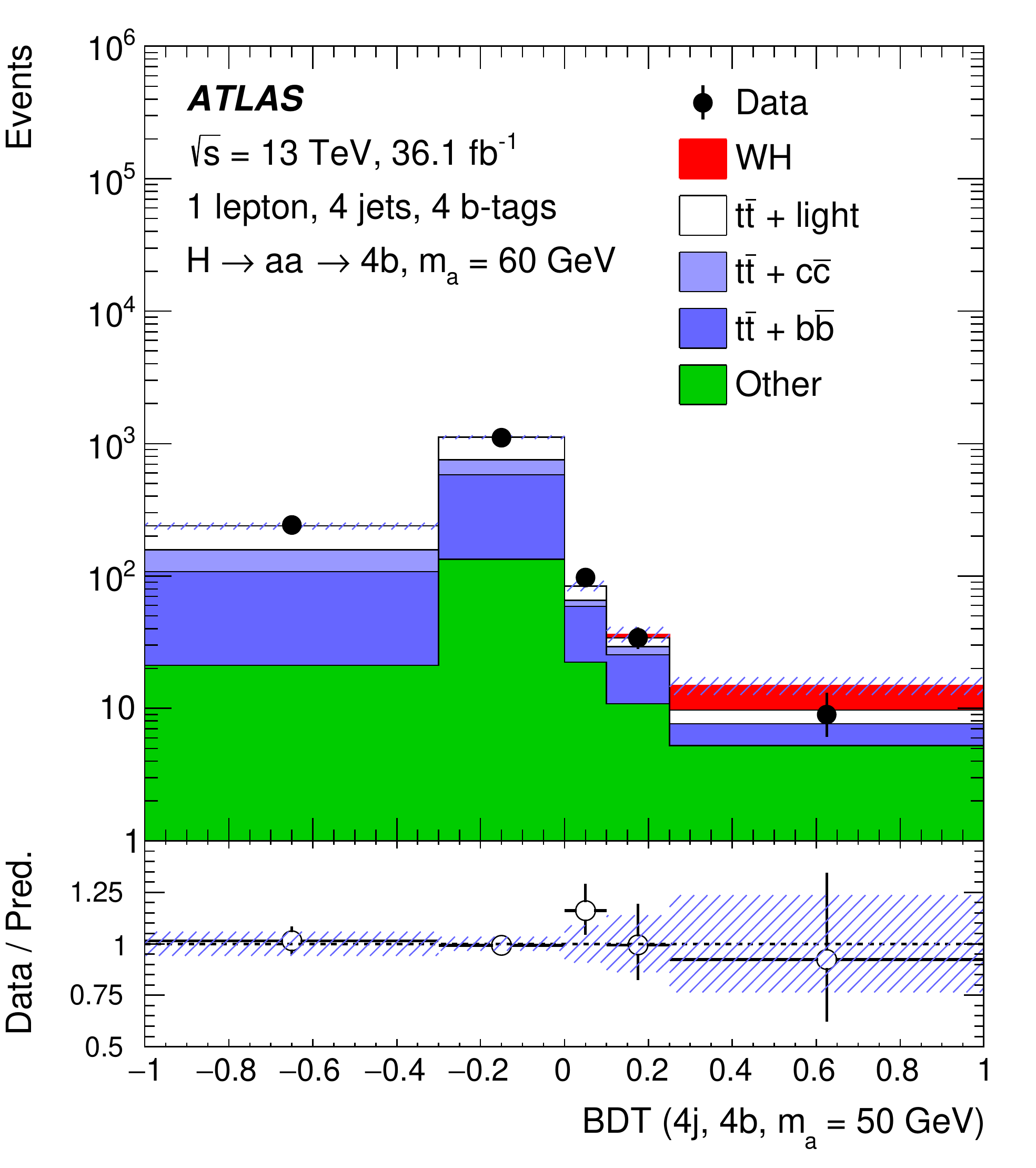}}%
\subfloat[]{\includegraphics[width=0.33\textwidth]{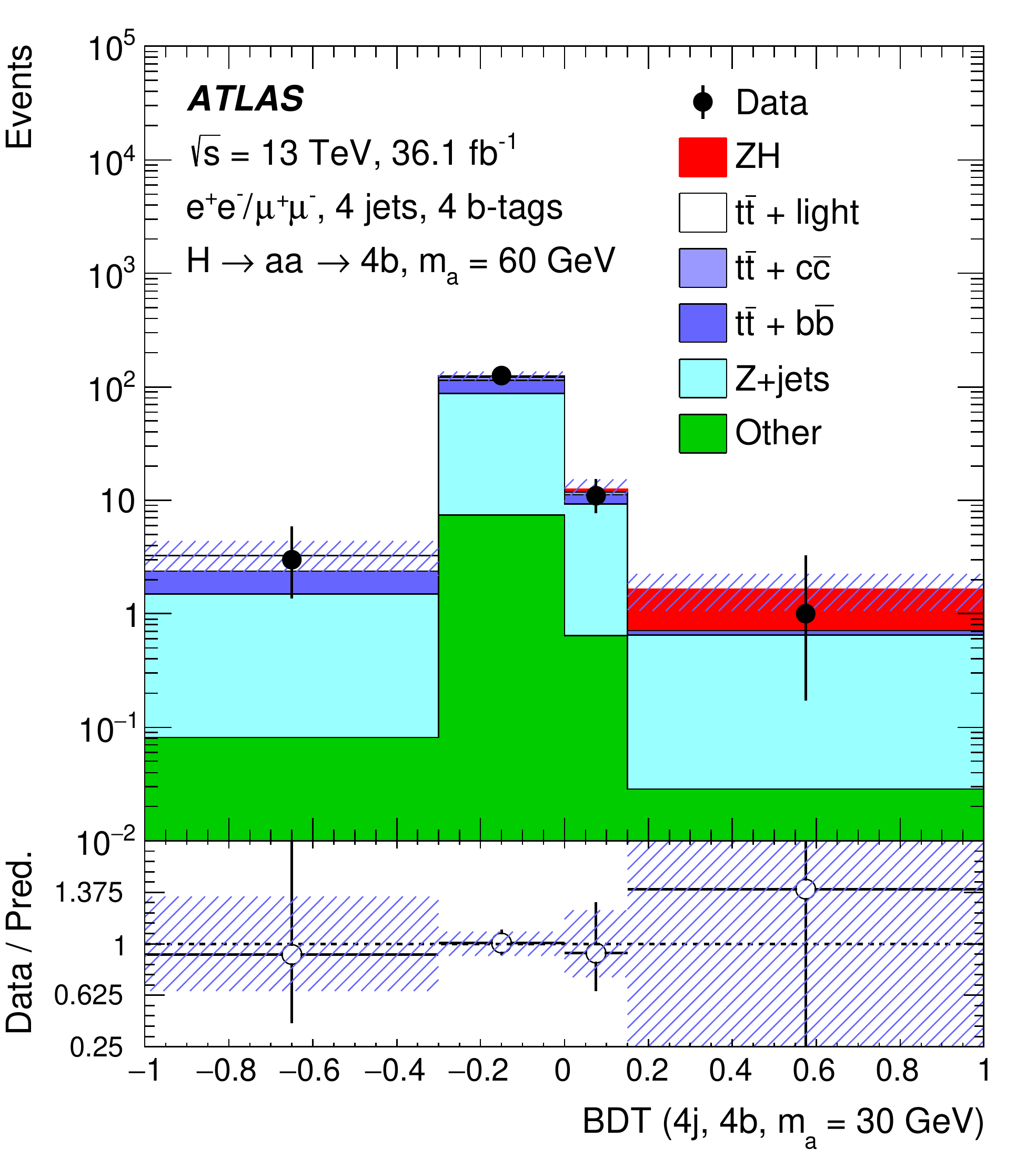}}\\
\caption{Comparison between data and prediction for the BDT discriminants
  in the (left) single-lepton signal regions trained with a signal
  with $m_a = 50~\GeV$: (a) (1$\ell$, 3j, 3b),
  (c) (1$\ell$, 4j, 3b) and (e) (1$\ell$, 4j, 4b) and in the (right)
  same-flavour dilepton signal regions trained with a signal with
  $m_a = 30~\GeV$: (b) (2$\ell$, 3j, 3b),
  (d) (2$\ell$, $\geq$4j, 3b) and (f) (2$\ell$, $\geq$4j,
  $\geq$4b), after performing the combined single-lepton and dilepton 
  fit of the predictions in all SRs and CRs to the data. The signal yield (solid red) is normalized to
  $\hat{\mu}$ after the fit for a signal with $m_a = 60~\GeV$. 
}
\label{fig:SR_fits}
\end{center}
\end{figure*}

Figures~\ref{fig:CR_sl} and~\ref{fig:CR_dl} show the distributions in
the control regions of the single-lepton and dilepton channels,
respectively, after performing the likelihood fit of these distributions to
data assuming $\mu=0$. The fit produces better agreement between the data
and the background predictions, and it reduces the uncertainties as a result
of the nuisance-parameter constraints and the correlations generated by the fit. 
Figure~\ref{fig:SR_fits} shows the post-fit background distributions for
the signal regions of the dilepton and single-lepton
channels. Table~\ref{tab:signalregions_postfityields} compares the observed event yield 
with the SM background prediction in each signal region, as well as the expected
number of signal events for a few representative values of
$m_a$. The overall acceptance times efficiency for signal is
approximately $0.2\%$ and $0.4\%$ for the signal regions (1$\ell$, 4j,
4b) and (2$\ell$, $\geq$4j, $\geq$4b), respectively. 

\begin{table}[h!]
\begin{center}
    \caption{\label{tab:signalregions_postfityields}Expected event
      yields of the post-fit background estimates and observed data in
      three signal regions for the single-lepton and dilepton
      channels after the combined single-lepton and dilepton fit to
      the data. The expected number of signal events for a
      few representative values of $m_{a}$ are indicated, assuming the
      SM $pp \rightarrow WH$ and $ZH$ cross-sections, $\sigma_{\mathrm{SM}}(WH) =
      1.37$~pb and $\sigma_{\mathrm{SM}}(ZH) =
      0.88$~pb~\cite{Heinemeyer:2013tqa}, respectively, and $\BR(H\rightarrow aa
      \rightarrow 4b) = 1$. The uncertainties include statistical and
      systematic components. The total uncertainty can differ from the
      sum in quadrature of individual sources due to correlations.
    }
\begin{small}
  \begin{tabular*}{\textwidth}{@{\extracolsep{\fill}}l|c c c | c c c}
\toprule
Process & (1$\ell$, 3j, 3b)  & (1$\ell$, 4j, 3b) & (1$\ell$, 4j, 4b) & (2$\ell$, 3j, 3b) & (2$\ell$, $\geq$4j, 3b) & (2$\ell$, $\geq$4j, $\geq$4b)\\
      \midrule
      $\ttlight$                    & \hd 9040 $\pm$ 490     & \hd 25600 $\pm$ 1300   & \hd 469 $\pm$ 76   & \hd 65 $\pm$ 25        & \hd 161 $\pm$ 48       & \hd 5 $\pm$ 4         \\
      $\ttcc$                 & \hd 1240 $\pm$ 570     & \hdd 4500 $\pm$ 1500    & \hdd 240 $\pm$ 100     & \hd 60 $\pm$ 32        & \hd 165 $\pm$ 66       & \hd 6 $\pm$ 6         \\
      $\ttbb$                 & \hd 1470 $\pm$ 690     & \hdd 4900 $\pm$ 1100    & \hdd 580 $\pm$ 150     & \hd 41 $\pm$ 16        & \hd 137 $\pm$ 43       & 30 $\pm$ 9        \\
      $\ttbar$ + $\gamma/\Wboson/\Zboson$  & \hd 18 $\pm$ 2         & \hd 88 $\pm$ 7         & \hd 16 $\pm$ 2        & \hd 1.2 $\pm$ 0.4      & \hd 45 $\pm$ 6         & \hd 7 $\pm$ 1         \\
      \Zjets{}                            & \hd 137 $\pm$ 75       & \hd 205 $\pm$ 67       & \hdd 8 $\pm$ 6         & 483 $\pm$ 42       & 1123 $\pm$ 73      & \hd 91 $\pm$ 17       \\
      \Wjets{}                            & \hd 1400 $\pm$ 380     & \hd 1550 $\pm$ 490     & \hdd 53 $\pm$ 20       & \hd - & \hdd - & \hd - \\
      Single top quark                      & \hd 1360 $\pm$ 160     & \hd 2210 $\pm$ 320     & \hdd 57 $\pm$ 31       & \hd 4 $\pm$ 2          & \hd 10 $\pm$ 3         & \hd 1 $\pm$ 1         \\
      Dibosons                            & \hdd - & \hdd - & \hdd - & 12 $\pm$ 7         & \hdd 45 $\pm$ 22        & \hd 1 $\pm$ 1\\
      Fakes                                 & \hdd 860 $\pm$ 280      & \hd 1130 $\pm$ 370     & \hdd 68 $\pm$ 20       & \hd 3 $\pm$ 1          & \hdd 8 $\pm$ 2          & \hd 0.3 $\pm$ 0.1     \\
      \hline
      Total                                 & 15520 $\pm$ 470    & 40190 $\pm$ 880    & 1485 $\pm$ 88     & 670 $\pm$ 31       & 1693 $\pm$ 60      & 140 $\pm$ 12      \\
      \hline
      Data                                  & 15428 \hddddd         & 40045 \hddddd             & 1488 \hdddd             & 670 \hdddd               & 1689  \hdddd             & 140   \hdddd            \\
      \hline
      $H \to aa \to 4b$               & \multicolumn{3}{c|}{$WH$} & \multicolumn{3}{c}{$ZH$} \\
      $m_{a} = 60$~\GeV{}                   & \hd 103 $\pm$ 24         & \hd 100 $\pm$ 21         & \hd 27 $\pm$ 9          & 28 $\pm$ 6          & \hdd 55 $\pm$ 10        & 12 $\pm$ 4         \\
      $m_{a} = 40$~\GeV{}                   & \hd 120 $\pm$ 25         & \hd 100 $\pm$ 19         & \hd 24 $\pm$ 7          & 32 $\pm$ 7          & \hdd 55 $\pm$ 10        & 11 $\pm$ 3         \\
      $m_{a} = 20$~\GeV{}                   & \hdd 65 $\pm$ 15          & \hdd 52 $\pm$ 11          & \hdd 9 $\pm$ 3           & 18 $\pm$ 4          & \hd 25 $\pm$ 5         & \hd 4 $\pm$ 1          \\
\bottomrule
      \end{tabular*}
  \end{small}
  \end{center}
  \end{table}

\begin{figure*}[htbp]
\begin{center}
\subfloat[]{\includegraphics[width=0.5\textwidth]{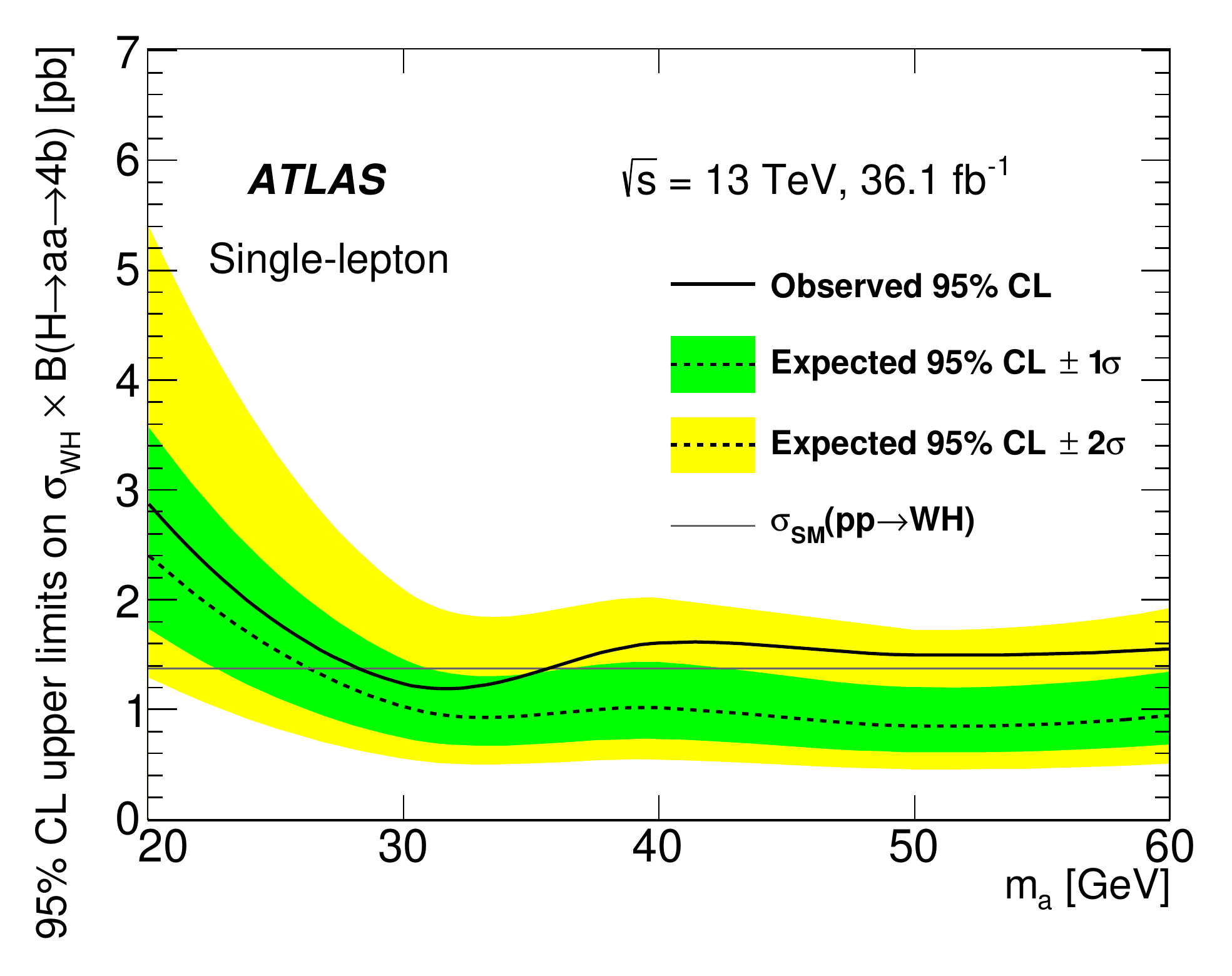}} %
\subfloat[]{\includegraphics[width=0.5\textwidth]{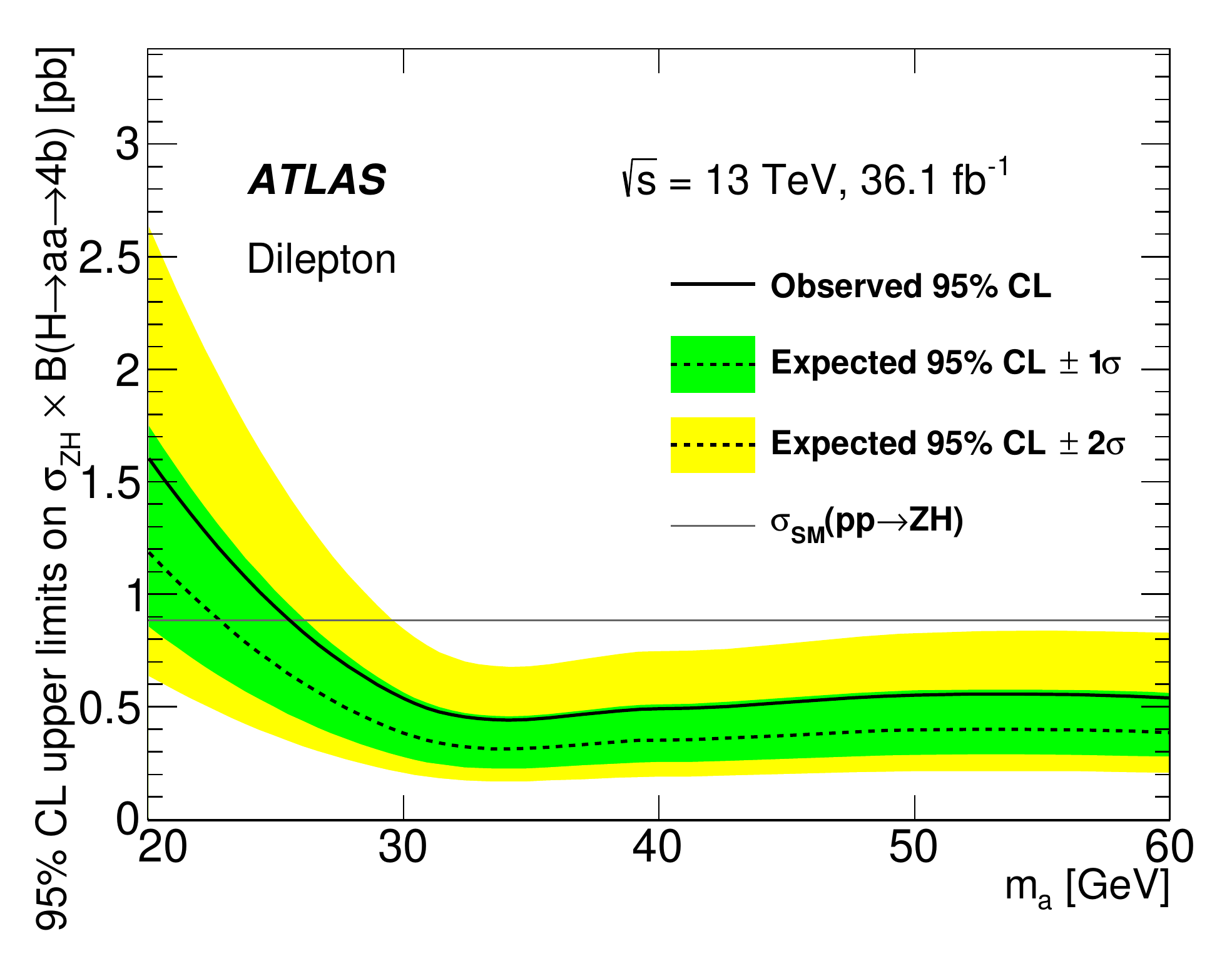}} \\
\includegraphics[width=0.5\textwidth]{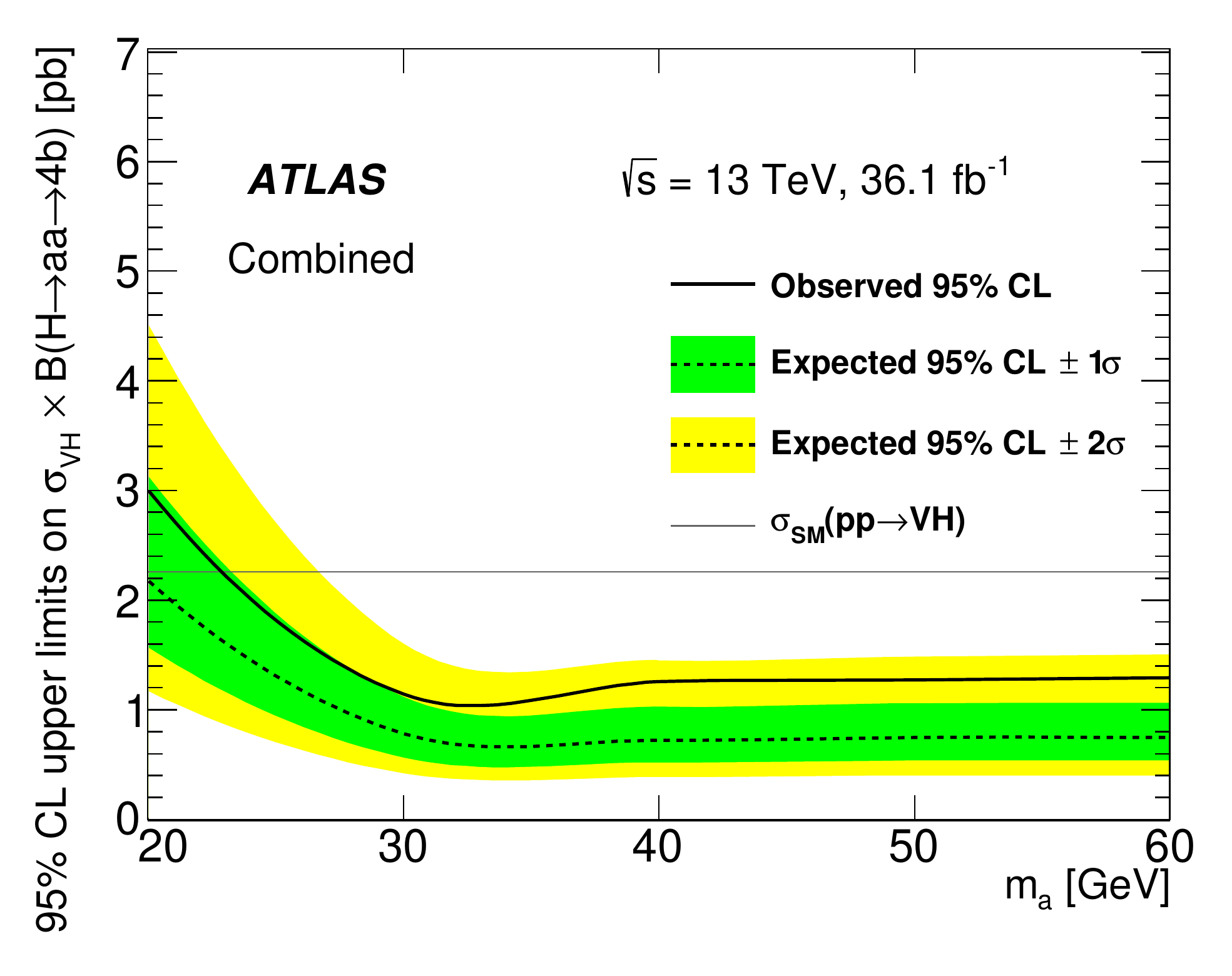} \\
\caption{Summary of the 95\% CL upper limits on
  $\sigma \times \BR(H\rightarrow aa \rightarrow 4b)$ 
  for (a) the single-lepton channel and (b) the dilepton channel, and
  (c) the combination of both channels.
The observed limits are shown, together with the expected limits
(dotted black lines). In the case of the expected limits, one- and
two-standard-deviation uncertainty bands are also displayed.}
\label{fig:limitPlot} 
\end{center}
\end{figure*}

No significant deviations from the SM predictions are found and
upper limits are set on $\sigma \times \BR(H\rightarrow aa \rightarrow
4b)$ for a mass hypothesis $m_a$ tested in steps of
$10~\GeV$ between $20$ and $60~\GeV$, as shown in
figure~\ref{fig:limitPlot}. The observed (expected) upper limits on $\mu$ for the
single-lepton channel range from 2.9 (2.4)~pb, for $m_a = 20~\GeV$, to 1.6
(0.94)~pb, for $m_a = 60~\GeV$. For the dilepton channel they range
from 1.6 (1.2)~pb, for $m_a = 20~\GeV$, to 0.54 (0.39)~pb for $m_a =
60~\GeV$. For comparison, the SM NNLO
cross-sections for $pp \rightarrow WH$ and $ZH$ are
$\sigma_{\mathrm{SM}}(WH) = 1.37$~pb and $\sigma_{\mathrm{SM}}(ZH)
=0.88$~pb, respectively~\cite{Heinemeyer:2013tqa}.
Upper limits are derived for the combination of the single-lepton and
dilepton channels assuming that the ratio of $WH$ to $ZH$
cross-sections follows the SM prediction. The observed 
(expected) 95\% CL upper limits on $\mu$ range from 3.0 (2.2)~pb, for
$m_a = 20~\GeV$, to 1.3 (0.74)~pb, for $m_a = 60~\GeV$. The combined
result excludes branching ratios as low as 0.45 in the mass range
$m_a$ between about 25~\GeV\ and 60~\GeV, assuming the
SM $WH$ and $ZH$ cross-sections. The reduced sensitivity for the
lighter $a$-boson hypotheses, where it is not possible to set limits
on the branching ratio, is due to a lower acceptance caused by
overlapping $b$-jets. 

\begin{figure*}[htbp]
\begin{center}
\includegraphics[width=0.5\textwidth]{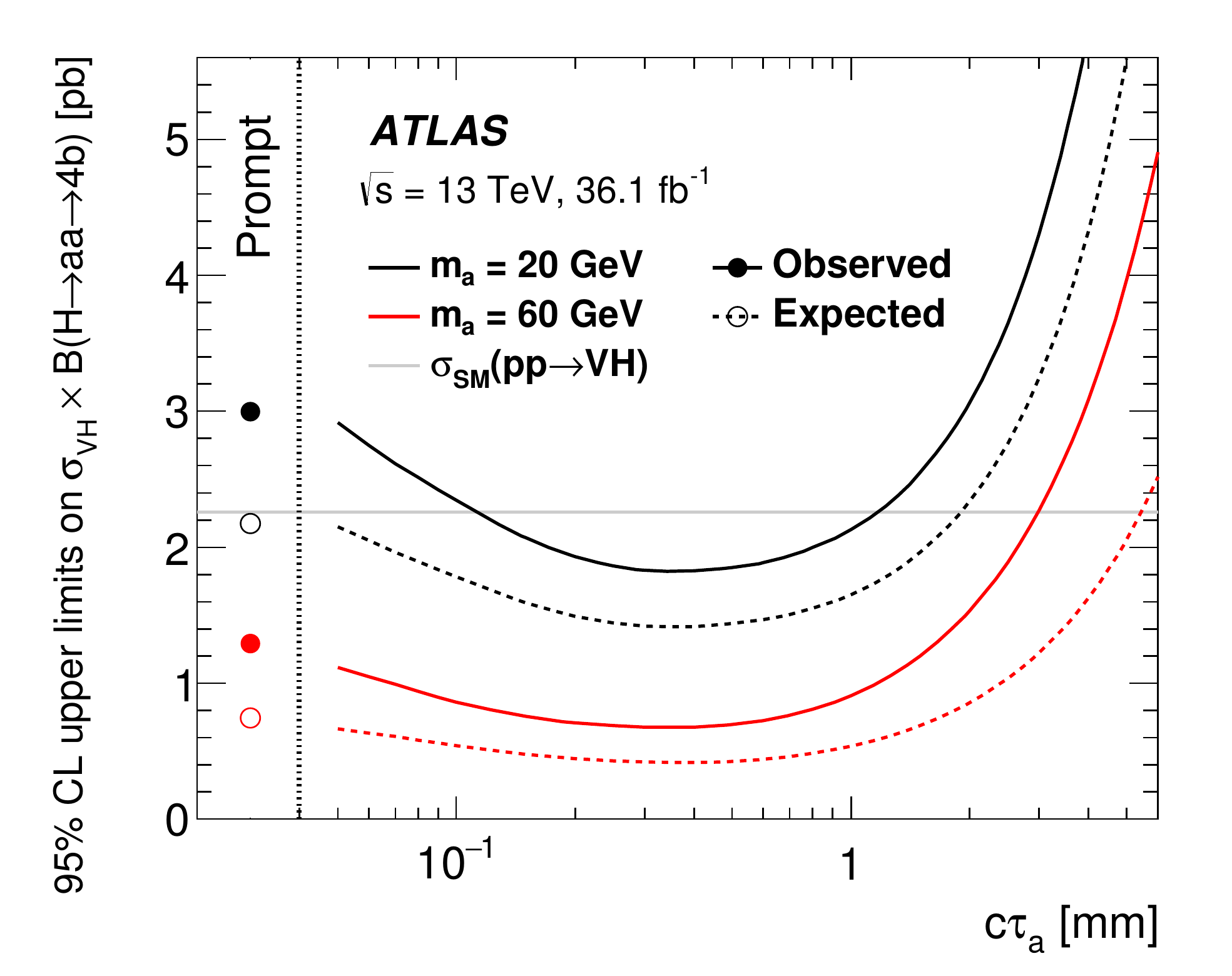}\\
\caption{Summary of the expected and observed combined 95\% CL upper
  limits on $\sigma \times \BR(H\rightarrow aa \rightarrow 4b)$ with
  prompt and long-lived $a$-bosons.  Expected (observed) limits for
  prompt $a$-bosons are shown as empty (filled) circles. Expected
  (observed) limits for long-lived signals are shown as dashed (solid) lines for two signal scenarios: $m_a =20$~and~$60$~\GeV.
}
\label{fig:limitPlotLL} 
\end{center}
\end{figure*}

Lifetime-dependent 95\% CL upper limits on the cross-section times 
$\BR(H\rightarrow aa \rightarrow 4b)$ for the combination of
single-lepton and dilepton channels are shown in
figure~\ref{fig:limitPlotLL}. For $a$-boson mean proper lifetimes up
to $\sim\! 0.5$~mm, the increased displacement of the $b$-jets leads
to an enhanced $b$-tagging efficiency, and the limits improve
slightly.  For longer $a$-boson lifetimes, the $b$-jets become
sufficiently displaced such that the jet reconstruction and
$b$-tagging efficiencies are diminished and the limits decrease rapidly.

\FloatBarrier

\section{Conclusion}
\label{sec:conclusion}

This paper presents a search for exotic decays of the Higgs boson into a
pair of spin-zero particles, $H \rightarrow aa$, where the
$a$-boson decays into two $b$-quarks. The search focuses on processes in
which the Higgs boson is produced in association with a $\Wboson$ or
$\Zboson$ boson that decays leptonically. The analysis uses the $pp$
collision dataset at $\sqrt{s} = 13~\TeV$ recorded by the ATLAS
detector at the LHC in 2015 and 2016, corresponding to an integrated
luminosity of $36.1 \pm 0.8$ fb$^{-1}$.
The search for $H \to aa \to 4b$ is performed in the mass range 
$20~\GeV \leq m_a \leq 60~\GeV$, considering prompt as well as 
long-lived $a$-bosons with mean proper lifetimes $c\tau_a$ up to 6~mm. The
analysis uses several kinematic variables combined in a multivariate
discriminant in signal regions and uses control regions to reduce 
uncertainties in the  background estimates. 
No significant excess of data is
observed relative to the SM predictions. Upper limits at 95\% CL are
derived for the $WH$ and $ZH$ production cross-section times branching
ratio of the decay $H \to aa \to 4b$. The combined observed upper limit
for promptly decaying $a$-bosons ranges from 3.0~pb for
$m_a= 20~{\GeV}$ to 1.3~pb for $m_a = 60~{\GeV}$, assuming that the ratio of $WH$ to $ZH$
cross-sections follows the SM prediction. For $a$-bosons with
longer proper lifetimes, the best limits are 1.8~pb and 0.68~pb, for
$m_a= 20~{\GeV}$ and $60~{\GeV}$ respectively, at $c\tau_a \sim 0.4$~mm.

\part*{Acknowledgements}
% Acknowledgements for papers with collision data
% Version 14-Feb-2018

% Standard acknowledgements start here
%----------------------------------------------
We thank CERN for the very successful operation of the LHC, as well as the
support staff from our institutions without whom ATLAS could not be
operated efficiently.

We acknowledge the support of ANPCyT, Argentina; YerPhI, Armenia; ARC, Australia; BMWFW and FWF, Austria; ANAS, Azerbaijan; SSTC, Belarus; CNPq and FAPESP, Brazil; NSERC, NRC and CFI, Canada; CERN; CONICYT, Chile; CAS, MOST and NSFC, China; COLCIENCIAS, Colombia; MSMT CR, MPO CR and VSC CR, Czech Republic; DNRF and DNSRC, Denmark; IN2P3-CNRS, CEA-DRF/IRFU, France; SRNSFG, Georgia; BMBF, HGF, and MPG, Germany; GSRT, Greece; RGC, Hong Kong SAR, China; ISF, I-CORE and Benoziyo Center, Israel; INFN, Italy; MEXT and JSPS, Japan; CNRST, Morocco; NWO, Netherlands; RCN, Norway; MNiSW and NCN, Poland; FCT, Portugal; MNE/IFA, Romania; MES of Russia and NRC KI, Russian Federation; JINR; MESTD, Serbia; MSSR, Slovakia; ARRS and MIZ\v{S}, Slovenia; DST/NRF, South Africa; MINECO, Spain; SRC and Wallenberg Foundation, Sweden; SERI, SNSF and Cantons of Bern and Geneva, Switzerland; MOST, Taiwan; TAEK, Turkey; STFC, United Kingdom; DOE and NSF, United States of America. In addition, individual groups and members have received support from BCKDF, the Canada Council, CANARIE, CRC, Compute Canada, FQRNT, and the Ontario Innovation Trust, Canada; EPLANET, ERC, ERDF, FP7, Horizon 2020 and Marie Sk{\l}odowska-Curie Actions, European Union; Investissements d'Avenir Labex and Idex, ANR, R{\'e}gion Auvergne and Fondation Partager le Savoir, France; DFG and AvH Foundation, Germany; Herakleitos, Thales and Aristeia programmes co-financed by EU-ESF and the Greek NSRF; BSF, GIF and Minerva, Israel; BRF, Norway; CERCA Programme Generalitat de Catalunya, Generalitat Valenciana, Spain; the Royal Society and Leverhulme Trust, United Kingdom.

The crucial computing support from all WLCG partners is acknowledged gratefully, in particular from CERN, the ATLAS Tier-1 facilities at TRIUMF (Canada), NDGF (Denmark, Norway, Sweden), CC-IN2P3 (France), KIT/GridKA (Germany), INFN-CNAF (Italy), NL-T1 (Netherlands), PIC (Spain), ASGC (Taiwan), RAL (UK) and BNL (USA), the Tier-2 facilities worldwide and large non-WLCG resource providers. Major contributors of computing resources are listed in Ref.~\cite{ATL-GEN-PUB-2016-002}.
%----------------------------------------------

\clearpage
\printbibliography

\clearpage
% ATLAS Collaboration author list
% Reference date of HIGG-2017-05 is 2018-03-29
% Author list last updated on date 03-SEP-18
% Data extracted on 03-Sep-2018 for paper reference HIGG-2017-05
% at 10:46am
 
\begin{flushleft}
{\Large The ATLAS Collaboration}

\bigskip

M.~Aaboud$^\textrm{\scriptsize 34d}$,    
G.~Aad$^\textrm{\scriptsize 99}$,    
B.~Abbott$^\textrm{\scriptsize 124}$,    
O.~Abdinov$^\textrm{\scriptsize 13,*}$,    
B.~Abeloos$^\textrm{\scriptsize 128}$,    
D.K.~Abhayasinghe$^\textrm{\scriptsize 91}$,    
S.H.~Abidi$^\textrm{\scriptsize 164}$,    
O.S.~AbouZeid$^\textrm{\scriptsize 39}$,    
N.L.~Abraham$^\textrm{\scriptsize 153}$,    
H.~Abramowicz$^\textrm{\scriptsize 158}$
% ...
\bigskip
\\

% ...
$^{13}$Institute of Physics, Azerbaijan Academy of Sciences, Baku; Azerbaijan.\\
% ...
$^{34}$$^{(a)}$Facult\'e des Sciences Ain Chock, R\'eseau Universitaire de Physique des Hautes Energies - Universit\'e Hassan II, Casablanca;$^{(b)}$Centre National de l'Energie des Sciences Techniques Nucleaires (CNESTEN), Rabat;$^{(c)}$Facult\'e des Sciences Semlalia, Universit\'e Cadi Ayyad, LPHEA-Marrakech;$^{(d)}$Facult\'e des Sciences, Universit\'e Mohamed Premier and LPTPM, Oujda;$^{(e)}$Facult\'e des sciences, Universit\'e Mohammed V, Rabat; Morocco.\\
% ...
$^{39}$Niels Bohr Institute, University of Copenhagen, Copenhagen; Denmark.\\
% ...
$^{91}$Department of Physics, Royal Holloway University of London, Egham; United Kingdom.\\
% ...
$^{99}$CPPM, Aix-Marseille Universit\'e, CNRS/IN2P3, Marseille; France.\\
% ...
$^{124}$Homer L. Dodge Department of Physics and Astronomy, University of Oklahoma, Norman OK; United States of America.\\
% ...
$^{128}$LAL, Universit\'e Paris-Sud, CNRS/IN2P3, Universit\'e Paris-Saclay, Orsay; France.\\
% ...
$^{153}$Department of Physics and Astronomy, University of Sussex, Brighton; United Kingdom.\\
% ...
$^{158}$Raymond and Beverly Sackler School of Physics and Astronomy, Tel Aviv University, Tel Aviv; Israel.\\
% ...
$^{164}$Department of Physics, University of Toronto, Toronto ON; Canada.\\
% ...
$^{a}$ Also at Borough of Manhattan Community College, City University of New York, NY; United States of America.\\
$^{b}$ Also at California State University, East Bay; United States of America.\\
$^{c}$ Also at Centre for High Performance Computing, CSIR Campus, Rosebank, Cape Town; South Africa.\\
$^{d}$ Also at CERN, Geneva; Switzerland.\\
$^{e}$ Also at CPPM, Aix-Marseille Universit\'e, CNRS/IN2P3, Marseille; France.\\
% ...
$^{*}$ Deceased

\end{flushleft}

% Created with Glance <Atlas.Glance@cern.ch>

\end{document}